\documentclass[11pt]{article}
\usepackage{amsmath,amsfonts,amssymb,amstext,latexsym,theorem}
\def\Id{{\rm 1\kern-.28em I}}
\def\D{{\rm /\kern-.65em {\cal D}}}
\def\R{{\rm I\kern-.2em R}}
\def\Z{{\rm Z\kern-.34em Z}}
\def\G{{\Gamma}}
\def\E{{\eta}}
\begin{document}
\title{\huge\bf  Supermatrix models for {\cal M}-theory \\based on $\mathfrak{osp}(1|32,\R)$}

\author{\bf Maxime Bagnoud$^1$, Luca Carlevaro$^1$ and Adel Bilal$^{1,2}$}
\date{}
\maketitle
\begin{center}
$^1${\it Institut de Physique, Universit\'e de Neuch\^atel \\
CH--2000 Neuch\^atel, Switzerland}

\vskip 5.mm

$^2$ {\it Laboratoire de Physique Th\'eorique, Ecole Normale Sup\'erieure\\
CNRS, UMR 8549 \\
24 rue Lhomond, 75231 Paris Cedex 05, France}

\end{center}

\vspace{1cm}

\begin{center}
{\bf Abstract}
\end{center}
\begin{quote}
Taking seriously the hypothesis that the full symmetry algebra of M-theory is $\mathfrak{osp}(1|32,\mathbb{R})$,
we derive the supersymmetry transformations for all fields that appear in 11- and 
12-dimensional realizations and give the associated {\cal SUSY} algebras. 
We study the background-independent $\mathfrak{osp}(1|32,\mathbb{R})$ cubic matrix model 
action expressed in terms of representations of the Lorentz groups $SO(10,2)$ and $SO(10,1)$.
We explore further the 11-dimensional case and compute an effective action for the BFSS-like degrees of freedom. 
We find the usual BFSS action with additional terms incorporating couplings to transverse 5-branes, 
as well as a mass-term and an infinite tower of higher-order interactions.
\end{quote}

\vspace{-170mm}
\begin{flushright}
NEIP--02-001 \\
LPTENS-02/10 \\
hep--th/0201183 \\
January 2002
\end{flushright}
\vspace{2.4cm}
\vfill
\begin{flushleft}
\rule{8.1cm}{0.2mm}\\

{\small e-mail: \tt Maxime.Bagnoud@, Luca.Carlevaro@, Adel.Bilal@unine.ch}\\
{\small Work supported in part by the Swiss Office for Education and Science,
the Swiss National Science Foundation and the European Community's Human 
Potential Programme under contract HPRN-CT-2000-00131 Quantum Spacetime.}\\
\end{flushleft}

\newpage
\pagenumbering{arabic}  
\section{Introduction}
M-theory \cite{Witten} should eventually provide a unifying framework for non-perturbative string theory. 
While there is lot of compelling evidence for this underlying M-theory, it is still a rather elusive theory, 
lacking a satisfactory intrinsic formulation. It is probably the matrix model by Banks, Fischler, Shenker and 
Susskind (BFSS) \cite{BFSS} which still comes closest to this goal.
In the absence of a microscopic description, quite some information can be obtained by simply looking at 
the eleven-dimensional superalgebra \cite{Townsend} whose central charges correspond to the extended objects,
i.e. membranes and five-branes present in M-theory. Relations with the hidden symmetries of eleven-dimensional 
supergravity \cite{CJS} and its compactifications and associated BPS configurations (see e.g. \cite{deWit,West}
and references therein) underlined further the importance of the algebraic aspects. It has been conjectured \cite{BVP} 
that the large superalgebra $\mathfrak{osp}(1|32)$ may play an important and maybe 
unifying r\^ole in M and F theory \cite{Vafa}.

The hidden symmetries of the 11$D$ supergravity action points to a non-linearly realized Lorentzian Kac-Moody algebra
$\mathfrak{e}_{11}$, whose supersymmetric extension contains $\mathfrak{osp}(1|32)$ as a finite-dimensional subalgebra. 
It would be interesting to investigate further the relationships between those two aspects of the symmetries underlying 
M-theory. 

In this paper, we have chosen to explore further the possible unifying r\^ole of $\mathfrak{osp}(1|32)$ and study 
its implications for matrix models. One of our main motivations is to investigate the dynamics of extended objects 
such as membranes and five-branes, when they are treated on the same footing as the ``elementary" degrees of freedom.
In order to see eleven and twelve-dimensional structures emerge, we have to embed the $SO(10,2)$ Lorentz algebra
and the $SO(10,1)$ Poincar\'e algebra into the large $\mathfrak{osp}(1|32)$ superalgebra. This will yield certain 
deformations and extensions of these algebras which nicely include new symmetry generators related to the charges 
of the extended objects appearing in the eleven and twelve-dimensional theories. The supersymmetry transformations 
of the associated fields also appear naturally. 

Besides these algebraic aspects, we are interested in the dynamics arising from matrix models derived from
such algebras. Following ideas initially advocated by Smolin~\cite{Smo1}, we start with matrices 
$M\in \mathfrak{osp}(1|32)$ as basic dynamical objects, write down a very simple action for them and then 
decompose the result according to the different representations of the eleven and twelve-dimensional algebras. 
In the eleven-dimensional case, we expect this action to contain the scalars $X_i$ of  
the BFSS matrix model and the associated fermions together with five-branes. 
In ten dimensions, cubic supermatrix models have already been studied by Azuma, Iso,  Kawai and Ohwashi \cite{AIKO} 
(more details can be found in Azuma's master thesis \cite{Azu}) in an attempt to compare it with the IIB
matrix model of Ishibashi, Kawai, Kitazawa and Tsuchiya \cite{IKKT}.

To test the relevance of our model, we try to exhibit its relations with the BFSS matrix model. For this purpose,
we perform a boost to the infinite momentum frame (IMF), thus reducing the explicit symmetry of the action to $SO(9)$.
Then, we integrate out conjugate momenta and auxiliary fields and calculate an effective action for the scalars 
$X_i$, the associated fermions, and higher form fields. What we obtain in the end is the BFSS matrix model with additional 
terms. In particular, our effective action explicitly contains couplings to 5-brane degrees of freedom, which are 
thus naturally incorporated in our model as fully dynamical entities. Moreover, we also get additional interactions 
and masslike terms. This should not be too surprising since we started with a larger theory.
The interaction terms we obtain are somewhat similar to the higher-dimensional operators one expects when integrating out 
(massive) fields in quantum field theory. This can be viewed as an extension of the BFSS theory
describing M-theoretical physics in certain non-Minkowskian backgrounds.

The outline of this paper is the following: in the next section we begin by recalling the form of 
the $\mathfrak{osp}(1|32)$ algebra and the decomposition of its matrices. In section 3 and 4, we study 
the embedding of the twelve-, resp. eleven-dimensional superalgebras into $\mathfrak{osp}(1|32)$, 
and obtain the corresponding algebraic structure including the extended objects described by a six- resp. five-form. 
We establish the supersymmetry transformations of the fields, and write down a cubic matrix model which 
yields an action for the various twelve- resp. eleven-dimensional fields. 
Finally, in section 5, we study further the eleven-dimensional matrix model, compute an effective action and do 
the comparison with the BFSS model.

\section{The $\mathfrak{osp}(1|32,\mathbb{R})$ superalgebra}
We first recall some definitions and properties of the unifying superalgebra $\mathfrak{osp}(1|32,\mathbb{R})$ 
which will be useful in the following chapters. The superalgebra is defined by the following three equations:  

\begin{align}\label{alg}
[Z_{AB},Z_{CD}] & = \Omega_{AD} Z_{CB} +  \Omega_{AC} Z_{DB} +  \Omega_{BD} Z_{CA} + \Omega_{BC} Z_{DA} \;, \notag\\
[Z_{AB},Q_C] & = \Omega_{AC} Q_B + \Omega_{BC} Q_A \;,\\
\{Q_A,Q_B\} & = Z_{AB}\;, \notag
\end{align}
where $\Omega_{AB}$ is the antisymmetric matrix defining the $\mathfrak{sp}(32,\mathbb{R})$ symplectic Lie algebra. 
Let us now give an equivalent description of elements of $\mathfrak{osp}(1|32,\mathbb{R})$. Following  Cornwell 
\cite{GTP3}, we call $\mathbb{R}B_L$ the real Grassmann algebra with 
$L$ generators, and $\mathbb{R}B_{L0}$ and $\mathbb{R}B_{L1}$ its even and odd subspace respectively. 
Similarly, we define a ($p|q$) supermatrix to be even (degree $0$) if it can be written as:
$$
M=\begin{pmatrix}
A & B\\
F & D\\
\end{pmatrix}\;.
$$
where $A$ and $D$ are $p \times p$, resp. $q \times q$  matrices with entries in $\mathbb{R}B_{L0}$, while $B$ and $F$ are 
$p \times q$ (resp. $q \times p$) matrices, with entries in $\mathbb{R}B_{L1}$. On the other hand, odd supermatrices 
(degree 1) are characterized by 4 blocks with the opposite parities.

We define the supertranspose of a supermatrix $M$ as\footnote{We warn the reader that this is not the same convention 
as in \cite{Azu}.}:
$$
M^{ST}=\begin{pmatrix}
A^{\top} & (-1)^{deg(M)} F^{\top}\\
-(-1)^{deg(M)} B^{\top} & D^{\top} \\
\end{pmatrix}\;.
$$
If one chooses the orthosymplectic metric to be the following $33 \times 33$ matrix:
$$
G=\begin{pmatrix}
0 & -\Id_{16} & 0\\
\Id_{16} & 0 & 0\\
0 & 0 & i
\end{pmatrix}\;,
$$
(where the $i$ is chosen for later convenience to yield a hermitian action),
we can define the $\mathfrak{osp}(1|32,\mathbb{R})$ superalgebra as the algebra of ($32|1$) supermatrices M satisfying 
the equation:
$$M^{ST} \cdot G + (-1)^{deg(Z)} G \cdot M = 0 \;.$$
From this defining relation, it is easy to see that an even orthosymplectic matrix should be of the form:
\begin{equation}\label{mpsi}
M=\begin{pmatrix}
A & B & \Phi_1\\
F & -A^{\top} & \Phi_2\\
-i\Phi_2^{\top} & i\Phi_1^{\top} & 0
\end{pmatrix}=
\begin{pmatrix}
m & \Psi \\
-i \Psi^{\top} C & 0
\end{pmatrix}\;,
\end{equation}
where A,B and F are $16 \times 16$ matrices with entries in $\mathbb{R}B_{L0}$ and $\Psi=(\Phi_1,\Phi_2)^{\top}$ is 
a 32-components Majorana spinors with entries in $\mathbb{R}B_{L1}$. Furthermore, $B=B^{\top}$, $F=F^{\top}$ so that 
$m \in \mathfrak{sp}(32,\mathbb{R})$ and $C$ is the following 
$32\times 32$ matrix:
\begin{equation}\label{Cdef}
C=\begin{pmatrix}
0 & -\Id_{16} \\
\Id_{16} & 0 
\end{pmatrix} \;,
\end{equation}
and will turn out to act as the charge conjugation matrix later on.

Such a matrix in the Lie superalgebra $\mathfrak{osp}(1|32,\mathbb{R})$ can
also be regarded as a linear combination of the generators thereof, which
we decompose in a bosonic and a fermionic part as:
\begin{equation}
H = 
\begin{pmatrix}
h & 0 \\
0 & 0 
\end{pmatrix}+
\begin{pmatrix}
0 & \chi \\
-i \chi^{\top} C & 0 
\end{pmatrix}
=h^{AB} Z_{AB} + \chi^A Q_A 
\end{equation}
where $Z_{AB}$ and $Q_A$ are the same as in~(\ref{alg}).
An orthosymplectic transformation will then act as:
\begin{equation}
\delta_H^{(1)} = 
[H,\bullet] = h^{AB} [Z_{AB},\bullet ] + \chi^A [Q_A,\bullet] =
\delta_{h}^{(1)} + \delta_{\chi}^{(1)}\; .
\end{equation}
This notation allows us to compute the commutation relations of two
orthosymplectic transformations characterized by $H=(h,\chi)$ and 
$E=(e,\epsilon)$. Recalling that for Majorana fermions $\chi^{\top} C \epsilon =\epsilon^{\top} C \chi$, 
we can extract from 
$[\delta_H^{(1)},\delta_E^{(1)}]$ the commutation relation of two symplectic
transformations:
\begin{equation}
[\delta_h^{(1)},\delta_e^{(1)}]_A^{\phantom{A}B} =
\begin{pmatrix}
[h,e]_A^{\phantom{A}B}  & 0 \\
0 & 0 
\end{pmatrix} \;,
\end{equation} 
the commutation relation between a symplectic transformation and
a supersymmetry:
\begin{equation}\label{com:rs}
[\delta_h^{(1)},\delta_\chi^{(1)}]_A^{\phantom{A}B} =
\begin{pmatrix}
0 & h_A^{\phantom{A}D} \chi_D \\
i (\chi^{\top} C)^D h_D^{\phantom{D}B} & 0
\end{pmatrix} \;,
\end{equation}
and the commutator of two supersymmetries:
\begin{equation}\label{com:ss}
[\delta_\epsilon^{(1)},\delta_\chi^{(1)}]_A^{\phantom{A}B} =
\begin{pmatrix}
-i(\chi_A (\epsilon^{\top} C)^B - \epsilon_A
(\chi^{\top} C)^B) & 0 \\
0 &  0
\end{pmatrix} \;.
\end{equation}

\section{The 12-dimensional case}

In order to be embedded into $\mathfrak{osp}(1|32,\mathbb{R})$, a Lorentz algebra
must have a fermionic representation of 32 real components at most. The
biggest number of dimensions in which this is the case is 12, where Dirac
matrices are $64\times 64$. As this dimension is even, there exists a Weyl representation
of 32 complex components. We need furthermore a Majorana condition to make
them real. This depends of course on the signature of space-time and is
possible only for signatures $(10,2)$, $(6,6)$ and $(2,10)$, when $(s,t)$ are
such that $s-t = 0 \mod 8$. Let us concentrate in this paper on the most
physical case (possibly relevant for F-theory) where the number of timelike dimensions is 2. 
However, since we choose to concentrate on the next section's M-theoretical case,
we will not push this analysis too far and will thus restrict ourselves to the computation 
of the algebra and the cubic action. 

To express the $\mathfrak{osp}(1|32,\mathbb{R})$ superalgebra in terms of
12-dimensional objects, we have to embed the $SO(10,2)$ Dirac matrices into
$\mathfrak{sp}(32,\mathbb{R})$ and replace the fundamental representation of
$\mathfrak{sp}(32,\mathbb{R})$ by $SO(10,2)$ Majorana-Weyl spinors. A
convenient choice of $64 \times 64$ Gamma matrices is the following: 
\begin{equation} 
\Gamma^0=\begin{pmatrix} 0 & -\Id_{32} \\ \Id_{32}
& 0 \\\end{pmatrix}  \text{  ,  } \Gamma^{11}=\begin{pmatrix} 0 &
\tilde{\Gamma}^0 \\ \tilde{\Gamma}^0 & 0 \\ \end{pmatrix} \text{  ,  }
\Gamma^i=\begin{pmatrix} 0 & \tilde{\Gamma}^i \\ \tilde{\Gamma}^i & 0 \\
\end{pmatrix}\phantom{a} \forall \, i = 1, \ldots, 10, \end{equation} where
$\tilde{\Gamma}^0$ is the $32 \times 32$ symplectic form: 
$$ \tilde{\Gamma}^0
= \begin{pmatrix} 0 & -\Id_{16} \\ \Id_{16} & 0 \\ \end{pmatrix}
$$
which, with the $\tilde{\Gamma}^i$'s and $\tilde{\Gamma}^{10}$, builds a Majorana representation of the 
$10+1$-dimensional Clifford algebra $\{\tilde{\Gamma}^\mu,\tilde{\Gamma}^\nu\} = 2
\eta^{\mu \nu} \Id_{32}$ for the mostly + metric
. Of course, $\tilde{\Gamma}^{10}= \tilde{\Gamma}^0 \tilde{\Gamma}^1 \ldots \tilde{\Gamma}^9$.
This choice has $(\Gamma^0)^2=(\Gamma^{11})^2=-\Id_{64}$, while $(\Gamma^i)^2=\Id_{64}$, 
$\forall i=1 \dots 10$, and gives a representation of $\{\Gamma^M,\Gamma^N\} = 2 \eta^{MN} \Id_{64}$ 
for a metric of the type $(-,+,\ldots,+,-)$. 
As we have chosen all $\Gamma$'s to be real, this allows to take $B=\Id$ in $\Psi^* = B \Psi$, which implies 
that the charge conjugation matrix $C=\Gamma^{0}\Gamma^{11}$, i.e.
$$
C=\begin{pmatrix}
-\tilde{\Gamma}^0 & 0 \\
0 & \tilde{\Gamma}^0 \\
\end{pmatrix}\;.
$$
This will then automatically satisfy:
\begin{equation}\label{CGam}
C \Gamma^M C^{-1} = (\Gamma^M)^{\top} \text{  ,  }
C \Gamma^{MN} C^{-1} = -(\Gamma^{MN})^{\top}
\end{equation}
and more generally:
\begin{equation}\label{CGamn}
C \Gamma^{M_1 \ldots M_n} C^{-1} = (-1)^{n(n-1)/2} (\Gamma^{M_1 \ldots M_n})^{\top}\; .
\end{equation}
The chirality matrix for this choice will be:
$$
\Gamma_* =  \Gamma^0 \ldots \Gamma^{11} = \begin{pmatrix}
-\Id_{32} & 0 \\
0 & \Id_{32} 
\end{pmatrix}\; .
$$
We will identify the fundamental representation of $\mathfrak{sp}(32,\mathbb{R})$ with positive 
chirality Majorana-Weyl spinors of $SO(10,2)$, i.e. those satisfying: ${\cal P}_+ \Psi = \Psi$, for:
$$
{\cal P}_+ = \frac{1}{2} (1 + \Gamma_*) =
\begin{pmatrix}
0 & 0 \\
0 & \Id_{32}
\end{pmatrix}\; .
$$
Decomposing the 64 real components of the positive chirality spinor $\Psi$ into $32+32$ or $16+16+16+16$, we can write:
$\Psi^{\top} = (0,\Phi^{\top}) = (0,0,\Phi_1^{\top}, \Phi_2^{\top})$. 
Because $\overline{\Psi}=\Psi^{\dagger} \Gamma^0 \Gamma^{11} = \Psi^{\top} C$,
this choice for the charge conjugation matrix $C$ is convenient since it will
act as $C$ in equation~(\ref{Cdef}) (though with a slight abuse of notation), and thus:
$$
(0,0,-i\Phi_2^{\top}, i\Phi_1^{\top}) = (0,-i \Phi^{\top} \tilde{\Gamma}^0) = -i \Psi^{\top} C = -i \overline{\Psi}.
$$

\subsection{Embedding of $SO(10,2)$ in $OSp(1|32,\mathbb{R})$}
We would now like to study how the Lie superalgebra of $OSp(1|32,\mathbb{R})$ 
can be expressed in terms of generators of the Super-Lorentz
algebra in 10+2 dimensions with additional symmetry generators. 
In other words, if we separate the $\mathfrak{sp}(32,\mathbb{R})$ 
transformations $h$ into a part sitting in the Lorentz algebra 
and a residual  $\mathfrak{sp}(32,\mathbb{R})$ part, we can give an
explicit description of this enhanced super-Poincar\'e algebra~(\ref{alg}) 
where we promote the former central charges to new generators of the
enhanced superalgebra.  

To do so, we need to expand a symplectic matrix in irreducible tensors 
of $SO(10,2)$. This can be done as follows:
\begin{equation}\label{h12}
h_A^{\phantom{A}B} = 
\frac{1}{2!} ({\cal P}_+ \Gamma^{MN})_A^{\phantom{A}B} h_{MN} + 
\frac{1}{6!} ({\cal P}_+ \Gamma^{M_1 \ldots M_6})_A^{\phantom{A}B} h^+_{M_1 \ldots M_6}
\end{equation}
where the $+$ on $h_{M_1 \ldots M_6}$ recalls its self-duality, and the components of $h$ in 
the decomposition in irreducible tensors of $SO(10,2)$ are given by  
$h_{MN} = -\frac{1}{32} Tr_{\mathfrak{sp}(32,\mathbb{R})}(h \Gamma_{MN})$ and 
$h^+_{M_1 \ldots M_6} = $ 
\linebreak $=-\frac{1}{32} Tr_{\mathfrak{sp}(32,\mathbb{R})}(h \Gamma_{M_1 \ldots M_6})$.
Indeed, a real symplectic $32 \times 32$ matrix satisfies $m \tilde{\Gamma}^0 = -\tilde{\Gamma}^0 m^{\top}$,
and $C$ acts like $\tilde{\Gamma}^0$ on ${\cal P}_{+} \Gamma^{M_1 \ldots M_n}$. Furthermore, (\ref{CGamn})
indicates that: 
\begin{equation}
C (1 + \Gamma_{*}) \Gamma^{M_1 \ldots M_n} = (-1)^{n(n-1)/2} ((1 + (-1)^n \Gamma_{*}) \Gamma^{M_1 \ldots M_n})^T C \;.
\end{equation} 
Thus, ${\cal P}_{+} \Gamma^{M_1 \ldots M_n}$ is symplectic iff $n$ is even and $(-1)^{n(n-1)/2}=-1$.
For $0 \leq n \leq 6$, this is only the case if $n=2$ or 6. As a matter of fact, the numbers of independent components
match since: $12\cdot 11/2 + 1/2 \cdot 12!/(6!)^2 =528 = 16 \cdot 33$.

The symplectic transformation $\delta_{h}$ may then be decomposed
into irreducible 12-dimensional tensors of symmetry generators, namely the $\mathfrak{so}(10,2)$  
Lorentz algebra generator $J^{MN}$  and a new 6-form symmetry generator $J^{M_1 \ldots M_6}$. 
To calculate the commutation relations of this enhanced Lorentz
algebra, we will choose the following representation of the symmetry generators:
$$
J^{MN} \,=\, \frac{1}{2!}\,{\cal P_+}\,\Gamma^{MN}\,,
\qquad J^{M_1 \ldots M_6} \,=\, \frac{1}{6!}\,{\cal P_+}\,\Gamma^{M_1 \ldots M_6}\;.
$$
so that a symplectic transformation will be given in this base by:
$$
h\,=\,h_{MN}\,J^{MN}\,+\,h_{M_1 \ldots M_6}\,J^{M_1 \ldots M_6}\;.
$$

We will now turn to computing the superalgebra induced by the above bosonic
generators and the supercharges for $D=10+2$. The bosonic commutators may readily be
computed using:
\begin{equation}\label{GamComm}
[\G_{M_{1}\ldots M_{k}},\G_{N_{1}\ldots N_{l}}]=\left\{
\begin{array}{l}\displaystyle{
\sum_{j=0}^{\lfloor (min(k,l)-1)/2\rfloor} \hspace{-.8cm}(-1)^{k-j-1}\,2\cdot (2j+1)! \binom{k}{2j+1} \binom{l}{2j+1}
\times}\\ 
\qquad\qquad\displaystyle{\times\E_{M_1 N_1}\ldots \E_{M_{2j+1} N_{2j+1}}\G_{M_{2j+2} N_{2j+2}\ldots M_{k} N_{l}}}
\text{ if $k\cdot l$ is even and,} \\ \\
\displaystyle{\sum_{j=0}^{(min(k,l)-1)/2} \hspace{-.75cm}(-1)^{j}\,2\cdot (2j)! \binom{k}{2j} \binom{l}{2j}\times}\\ 
\qquad\qquad\displaystyle{\times \E_{M_1 N_1}\ldots \E_{M_{2j} N_{2j}}\G_{M_{2j+1} N_{2j+1}\ldots M_{k} N_{l}}}\quad\quad \text{ if $k\cdot l$ is odd.}
\end{array}\right.
\end{equation} 
On the other hand,
for the commutation relations involving fermionic generators, we proceed as
follows. We expand equation (\ref{com:rs}) of the preceding chapter in irreducible tensors
of $SO(10,2)$:
\begin{align}
[\delta_{\chi},\delta_h]\,=&\,
-\frac{1}{2!}\,\chi^A\,h_{MN} ({\cal P_+}\,\Gamma^{MN})^B_{\phantom{B}A} Q_B-
\frac{1}{6!}\,\chi^A\,h_{M_1\ldots M_6} ({\cal P_+}\,\Gamma^{M_1\ldots M_6})^B_{\phantom{B}A} Q_B\;,\notag
\end{align}
which is also given by:
\begin{equation}
[\delta_{\chi},\delta_h]
\,=\,\chi^A\,h_{MN}[Q_A,J^{MN}]+\chi^A\,h_{M_1\ldots M_6}[Q_A,J^{M_1\ldots M_6}]\;.
\end{equation}
Comparing terms pairwise, we see that the supercharges transform as:
\begin{equation*}
[J^{MN},Q_A]\,=\,\frac{1}{2!}\, ({\cal P_+}\,\Gamma^{MN})^B_{\phantom{B}A} Q_B\;,\qquad
[J^{M_1\ldots M_6},Q_A]\,=\,\frac{1}{6!}\,({\cal P_+}\,J^{M_1\ldots M_6})^B_{\phantom{B}A} Q_B \;.
\end{equation*}
Finally, in order to obtain the anti-commutator of two supercharges, we expand the RHS of (\ref{com:ss}) in the 
bosonic generators $J^{MN}$ and $J^{M_1 \ldots M_6}$:
\begin{equation}
\label{QaQb} 
-\chi^A\epsilon_B \{Q_A,Q^B\}\,\equiv\,
[\delta_{\chi},\delta_{\epsilon}]\,=\,
\frac{i}{16} (\chi^{\top}C \Gamma_{MN}\epsilon) J^{MN}
+\frac{i}{16} (\chi^{\top}C \Gamma_{M_1\ldots M_6}\epsilon)
J^{M_1\ldots M_6}\;,
\end{equation}
and match the first and the last term of the equation.

Summarizing the results of this section, we get the following
12-dimensional realization of the superalgebra
$\mathfrak{osp}(1|32,\mathbb{R})$\footnote{Notice that the second term appearing on the right handside of the
third commutator is in fact proportional to $\Gamma^{M_1\ldots
M_{10}}$, which, in turn, can be reexpressed as 
$\Gamma^{M_1\ldots M_{10}}=(1/2)\varepsilon^{AB\,M_1\ldots
M_{10}}\,\Gamma_{AB} \Gamma_*$. Indeed, in $10+2$ dimensions, we
always have:
\begin{equation*}\label{eps}
\Gamma^{M_1\ldots M_{k}}=\frac{1}{(12-k)!}\varepsilon^{M_1\ldots
M_{k} M_{k+1} \ldots M_{12}}\,\Gamma_{M_{k+1} \ldots M_{12}} \Gamma_*
\end{equation*}}:
\begin{align}\label{algebra}
[J^{MN},J^{OP}] &\,=\, -4 \eta^{[M [O} J^{N] P]} \notag\\
[J^{MN},J^{M_1\ldots M_6}] &\,=\,-12\,\eta^{[M[M_1}\,J^{N]M_2\ldots M_6]} \notag\\ 
[J^{N_1\ldots N_6},J^{M_1\ldots M_6}] 
&\,=\,-4!\,6!\,\eta^{[N_1[M_1}\,\eta^{N_2\,M_2}\,\eta^{N_3\,M_3}\,\eta^{N_4\,M_4}\,\eta^{N_5\,M_5}\,J^{N_6]M_6]} \notag\\
&\,+\,2\cdot 6^2\,
\eta^{[N_1[M_1}\,\varepsilon^{N_2\ldots N_6]M_2\ldots M_6]}_{\phantom{N_2\ldots N_6]M_2\ldots M_6]}AB}\,J^{AB}\notag\\
&\,+\,4\left(\frac{6!}{4!}\right)^3\,\eta^{[N_1[M_1}\,\eta^{N_2\,M_2}\,\eta^{N_3\,M_3}\,J^{N_4\ldots N_6]M_4\ldots M_6]}
\end{align}
\begin{align*}
[J^{MN},Q_A]\,&=\,\frac{1}{2}\, ({\cal P_+}\,\Gamma^{MN})^B_{\phantom{B}A} Q_B\\
[J^{M_1\ldots M_6},Q_A]\,&=\,\frac{1}{6!}\,({\cal P_+}\,\Gamma^{M_1\ldots M_6})^B_{\phantom{B}A} Q_B\\
\{Q_A,Q^B\} \,&=\, -\frac{i}{16} (C \Gamma_{MN})_A^{\phantom{A}B} J^{MN}
- \frac{i}{16} (C \Gamma_{M_1\ldots M_6})_A^{\phantom{A}B}
J^{M_1\ldots M_6}\;, 
\end{align*}
where antisymmetrization brackets on the RHS are meant to match the anti-symmetry of indices on the LHS.

\subsection{Supersymmetry transformations of 12$D$ matrix fields}
In the following, we will construct a dynamical matrix model based on the symmetry group
$\mathfrak{osp}(1|32,\mathbb{R})$ using elements in the adjoint representation of
this superalgebra, i.e. matrices in this superalgebra. We can write such a matrix
as:
\begin{equation}
M\,=\,
\begin{pmatrix}
m & \Psi \\
-i \Psi^{\top} C & 0 
\end{pmatrix}\;,
\end{equation} 
where $m$ is in the adjoint representation of
$\mathfrak{sp}(32,\mathbb{R})$ and $\Psi$ is in the fundamental.
Since $M$ belongs to the adjoint representation, a SUSY will act on it
in the following way: 
\begin{equation}\label{delchi}
\delta^{(1)}_{\chi} M_A^{\phantom{A} B} = \chi^D [Q_D,M]_A^{\phantom{A} B} = 
\begin{pmatrix}
-i(\chi_A (\Psi^{\top} C)^B - \Psi_A (\chi^{\top} C)^B) & - m_A^{\phantom{A}D} \chi_D \\
-i (\chi^{\top} C)^D m_D^{\phantom{C}B} & 0
\end{pmatrix}
\end{equation}
In our particular 12D setting, $m$ gives rise to a 2-form field $C$ (with $SO(10,2)$ indices, 
not to be confused with the charge conjugation matrix with $\mathfrak{sp}(32,\mathbb{R})$ indices) 
and a self-dual 6-form field $Z^+$, as follows:
\begin{equation}\label{m12}
m_A^{\phantom{A}B} = 
\frac{1}{2!} ({\cal P}_+ \Gamma^{MN})_A^{\phantom{A}B} C_{MN} + 
\frac{1}{6!} ({\cal P}_+ \Gamma^{M_1 \ldots M_6})_A^{\phantom{A}B} Z^+_{M_1 \ldots M_6}\;.
\end{equation}
We can extract the supersymmetry transformations of $C$, $Z^+$ and $\Psi$ from~(\ref{delchi}) and we obtain:
\begin{align}\label{del6}
\delta_{\chi}^{(1)} C_{MN} &= \frac{i}{16} \overline{\chi}\,\Gamma_{MN} \Psi \;,\notag\\
\delta_{\chi}^{(1)} Z^+_{M_1 \ldots M_6} &= \frac{i}{16} \overline{\chi}\,\Gamma_{M_1 \ldots M_6} \Psi  \;,\\
\delta_{\chi}^{(1)} \Psi &= - \frac{1}{2}\Gamma^{MN} \chi\, C_{MN} -
\frac{1}{6!}\Gamma^{M_1 \ldots M_6} \chi\, Z^+_{M_1 \ldots M_6} \;.\notag
\end{align}
These formul\ae\; allow us to compute the effect of two successive supersymmetry
transformations using~(\ref{CGamn}) and~(\ref{GamComm}):
\begin{align}
\label{com:ssp}
[\delta_{\chi}^{(1)},\delta_{\epsilon}^{(1)}] \Psi 
&=\, \frac{i}{16} \Big\{( \overline{\epsilon} \Psi)\, \chi 
- (\overline{\chi}\, \Psi)\, \epsilon \Big\} \;,\notag  \\
&\notag \\
[\delta_{\chi}^{(1)},\delta_{\epsilon}^{(1)}] C_{MN} &= \,
\frac{i}{4}\overline{\chi}\, \Big\{\Gamma_{[M}^{\phantom{[M}P} C_{N]P} + 
\frac{1}{5!}\Gamma_{[M}^{\phantom{[M}M_1 \ldots M_5} Z^+_{N] M_1 \ldots M_5}\Big\}{\cal P}_+\, \epsilon\;,\
\end{align}
\begin{align}
[\delta_{\chi}^{(1)},\delta_{\epsilon}^{(1)}] Z^+_{M_1 \ldots M_6} &=\, 
\overline{\chi} \,\Big\{\frac{3i}{4} \Gamma_{[M_1 \ldots M_5}^{\phantom{[M_1 \ldots M_5}N} C_{M_6]N} 
+ \frac{3i}{2} \Gamma_{[M_1}^{\phantom{[M_1}N} Z^+_{M_2 \ldots M_6] N} -\notag\\ 
&- \frac{5i}{12} \Gamma_{[M_1 M_2 M_3}^{\phantom{[M_1 M_2 M_3}N_1 N_2 N_3} Z^+_{M_4 M_5 M_6] N_1 N_2 N_3} 
\Big\}{\cal P}_+\, \epsilon\;, \notag
\end{align}
where we used the self-duality\footnote{$Z^+$ satisfies $Z^+_{M_1 \ldots M_6} = 
\frac{1}{6!}\varepsilon_{M_1 \ldots M_6}^{\phantom{M_1 \ldots M_6}N_1 \ldots N_6}Z^+_{N_1 \ldots N_6}$} of $Z^+$.
At this stage, we can mention that the above results are in perfect agreement with the adjoint representation
of $[\delta_{\chi}^{(1)},\delta_{\epsilon}^{(1)}]$ (viz.~(\ref{com:ss})~) on the matrix fields. 

\subsection{$\mathfrak{sp}(32,\mathbb{R})$ transformations of the
fields and their commutation relation with supersymmetries}
To see under which transformations an $\mathfrak{osp}(1|32,\mathbb{R})$-based matrix model should be invariant,
one should look at the full transformation properties including the bosonic $\mathfrak{sp}(32,\mathbb{R})$ 
transformations. In close analogy with equation~(\ref{delchi}), we have the following full transformation law of $M$:
\begin{equation}\label{del}
\delta_H^{(1)}M_A^{\phantom{A}B}\, =\, \left[
\begin{pmatrix}
h & \chi \\
-i \overline{\chi} & 0 
\end{pmatrix}
,
\begin{pmatrix}
m & \Psi \\
-i \overline{\Psi} & 0 
\end{pmatrix}
\right]_A^{\phantom{A}B}, 
\end{equation}
implying the following transformation rules:
\begin{align}
\delta_{H}^{(1)} m_A^{\phantom{A}B} &= 
[h,m]_A^{\phantom{A}B} - i (\chi_A \overline{\Psi}^B - \Psi_A \overline{\chi}^B) \;,\\
\delta_{H}^{(1)} \Psi_A &= h_A^{\phantom{A}C} \Psi_C - m_A^{\phantom{A}C} \chi_C \;.
\end{align}
We then want to extract from the first of the above equations the full
transformation properties  of $C_{MN}$ and $Z^+_{M_1 \ldots M_6}$.
From~(\ref{algebra}) and~(\ref{com:ssp}) or directly using~(\ref{GamComm}) and the cyclicity of the trace, 
the bosonic transformations are: 
\begin{align}\label{deltah}
\delta_{h}^{(1)} C_{MN} &=\; 
\, 4 h^{P}_{\phantom{P}[N} C_{M]P} \,+\, 
\frac{4}{5!} h^{N_1 \ldots N_5}_{\phantom{N_1 \ldots N_5}[N} 
Z^+_{M]N_1 \ldots N_5} \;,\notag\\
\delta_{h}^{(1)} Z^+_{M_1 \ldots M_6} &=\;
\,12h_{[M_1 \ldots M_5}^{\phantom{[M_1 \ldots M_5}P} C_{M_6] P} 
\,-\,24 \, h^N_{\phantom{N}[M_1} Z^+_{M_2 \ldots M_6] N}\,-\\
&+\, \frac{20}{3} h^{N_1 N_2 N_3}_{\phantom{N_1 N_2 N_3}[M_1 M_2 M_3} 
Z^+_{M_4 M_5 M_6] N_1 N_2 N_3}\;,\notag
\end{align}
while the fermionic part is as in~(\ref{del6}). 
If one uses~(\ref{deltah}) to compute the commutator of a supersymmetry 
and an $\mathfrak{sp}(32,\mathbb{R})$ transformation, the results will look very complicated.
On the other hand, the commutator of two symmetry transformations may be cast in a compact
form using the graded Jacobi identity of the $\mathfrak{osp}(1|32,\mathbb{R})$ superalgebra, 
which comes into the game since matrix fields are in the adjoint representations of
$\mathfrak{osp}(1|32,\mathbb{R})$. 

Such a commutator acting on the fermionic field $\Psi$ yields: 
\begin{align} \label{compsi}
[\delta_{\chi}^{(1)},\delta_{h}^{(1)}] \,\Psi &=\; -h m \chi \,+\,
[h,m]\chi \;= \; -m h\chi \;=\notag \\
&= \;- \frac{1}{2!} ({\cal P}_+ \Gamma^{MN}h\chi) C_{MN} \,-\, 
\frac{1}{6!} ({\cal P}_+ \Gamma^{M_1 \ldots M_6}h\chi)Z^+_{M_1 \ldots M_6}\;. 
\end{align} 
The same transformation on $m$ leads to: 
\begin{equation}\label{comM}
[\delta_{\chi}^{(1)},\delta_{h}^{(1)}]\,m_A^{\phantom{A}B}\,=\,
i\left(\Psi_A(\chi^{\top}h^{\top}C)^B -(h\chi)_A(\Psi^{\top}C)^B\right)\;, 
\end{equation}
which in components reads:
\begin{align}\label{comC}
[\delta_{\chi}^{(1)},\delta_{h}^{(1)}] \,C_{MN}\,=\,\frac{i}{16}\,\chi^{\top}C h\Gamma_{MN}\Psi\;,\\
\label{comZ}
[\delta_{\chi}^{(1)},\delta_{h}^{(1)}]\, Z^+_{M_1 \ldots M_6} \,=\,
\frac{i}{16}\,\chi^{\top}C h\Gamma_{M_1 \ldots M_6}\Psi\;.
\end{align}
In eqns. (\ref{compsi}), (\ref{comC}) and (\ref{comZ}), one could write $h$ in components as in~(\ref{h12}) 
and use:
\begin{align}\label{GamProd}
\G_{M_{1}\ldots M_{k}}\G_{N_{1}\ldots N_{l}}=\sum_{j=0}^{min(k,l)} (-1)^{k-j-1}\,2\cdot j! \binom{k}{j} \binom{l}{j} 
\E_{M_1 N_1}\ldots \E_{M_{j} N_{j}}\G_{M_{j+1} N_{j+1}\ldots M_{k} N_{l}} 
\end{align}
to develop the products of Gamma matrices in irreducible tensors of $SO(10,2)$ and obtain 
a more explicit result. The final expression for (\ref{compsi}) and (\ref{comZ}) will contain Gamma matrices 
with an even number of indices ranging from 0 to 12, while in (\ref{comC}) the number of indices will stop at 8. 
Since we won't use this result as such in the following, we won't give it here explicitly.

\subsection{A note on translational invariance and kinematical supersymmetries} 
At this point, we want to make a comment on so-called kinematical supersymmetries that have been discussed
in the literature on matrix models (~\cite{IKKT},~\cite{AIKO}). Indeed, commutation relations of dynamical 
supersymmetries do not close to give space-time translations, i.e. they do not shift the target
space-time fields $X^M$ by a constant vector.

However, as was pointed out in~\cite{IKKT} and~\cite{AIKO}, if one introduces so-called kinematical supersymmetry 
transformations, their commutator with dynamical supersymmetries yields the expected translations by a constant vector. 
By kinematical supersymmetries, one simply means translations of fermions by a constant Grassmannian odd parameter. 
In our case, this assumes the form:  
\begin{align}\label{trans}  
\delta_{\xi}^{(2)} C_{MN} \,=\, \delta_{\xi}^{(2)} Z^+_{M_1 \ldots M_6} &\,=\, 0\;, \qquad\qquad
\delta_{\xi}^{(2)} \Psi
\,= \,\xi\;,
 \\ \Longrightarrow
[\delta_{\xi}^{(2)},\delta_{\zeta}^{(2)}] M &\,=\, 0 \notag
\end{align}
Since there is no vector field to be interpreted as space-time coordinates in this 12-dimensional 
setting, it is interesting to look at the interplay between dynamical and kinematical supersymmetries 
(which we denote respectively by $\delta^{(1)}$ and $\delta^{(2)}$) when acting on higher-rank tensors.
In our case:
\begin{equation}\label{eq:trans}
[\delta_{\chi}^{(1)},\delta_{\xi}^{(2)}] C_{MN} \,=\, -\frac{i}{16} (\chi^{\top} C \Gamma_{MN} \xi)\;,\qquad
[\delta_{\chi}^{(1)},\delta_{\xi}^{(2)}] Z^+_{M_1 \ldots M_6} = -\frac{i}{16}
(\chi^{\top} C \Gamma_{M_1 \ldots M_6} \xi)\;.
\end{equation} 
Thus, $[\delta_{\chi}^{(1)},\delta_{\xi}^{(2)}]$ applied to $p$--forms closes
to translations by a constant $p$-form, generalizing the vector case mentioned
above.

For fermions, we have as expected:
\begin{equation} 
[\delta_{\chi}^{(1)},\delta_{\xi}^{(2)}] \Psi = 0\;.
\end{equation} 

It is however more natural to consider dynamical and kinematical symmetries to be independent.
We would thus expect them to commute. With this in mind, we suggest a generalised version of 
the translational symmetries introduced in~(\ref{trans}):
\begin{equation}
\delta_{K}^{(2)} \Psi = \xi, \phantom{ab}
\delta_{K}^{(2)} C_{MN} = k_{MN}, \phantom{ab}
\delta_{K}^{(2)} Z^+_{M_1 \ldots M_6} = k^+_{M_1 \ldots M_6}\;.
\end{equation}
It is then natural that the matrix
\begin{equation}
K=\begin{pmatrix}
k & \xi \\
-i \xi^{\top} C & 0
\end{pmatrix}
\end{equation}
should transform in the adjoint of $\mathfrak{osp}(1|32,\mathbb{R})$, which means that:
\begin{align}
\delta_{H}^{(1)} k_A^{\phantom{A}B} &= 
[h,k]_A^{\phantom{A}B} - i (\chi_A (\xi^{\top} C)^B - \xi_A (\chi^{\top} C)^B) \\
\delta_{H}^{(1)} \xi_A &= h_A^{\phantom{A}C} \xi_C - k_A^{\phantom{A}C} \chi_C. 
\end{align}
We can now compute the general commutation relations between translational symmetries $M \rightarrow M\,+\,K$ and
$\mathfrak{osp}(1|32,\mathbb{R})$ transformations and conclude that these operations actually 
commute:
\begin{equation}
[\delta_{H}^{(1)},\delta_{K}^{(2)}] M = 0\;.
\end{equation}

\subsection{12-dimensional action for supersymmetric cubic matrix model} 
We will now build the simplest gauge- and translational-invariant 
$\mathfrak{osp}(1|32,\mathbb{R})$ supermatrix model with $U(N)$ gauge group. 
For this purpose, we promote each entry of the matrix $M$ to a hermitian matrix in the Lie algebra
of $\mathfrak{u}(N)$ for some value of $N$. We choose the generators $\{t^a\}_{a=1, \ldots , N^2}$ of 
$\mathfrak{u}(N)$ so that: $[t^a,t^b]=if^{abc}t^c$ and $Tr_{\mathfrak{u}(N)}(t^a \cdot t^b)=\delta^{ab}$.

In order to preserve both orthosymplectic and gauge invariance of the model, it suffices to write
its action as a supertrace over $\mathfrak{osp}(1|32,\mathbb{R})$ and a trace over $\mathfrak{u}(N)$ 
of a polynomial of $\mathfrak{osp}(1|32,\mathbb{R})\otimes\mathfrak{u}(N)$ matrices. 
Following~\cite{Smo1}, we consider the simplest model containing interactions, namely:\linebreak 
$STr_{\mathfrak{osp}(1|32,\mathbb{R})} Tr_{\mathfrak{u}(N)} (M[M,M]_{\mathfrak{u}(N)})$. 
For hermiticity's sake one has to multiply such an action by a factor of $i$.
We also introduce a coupling constant $g^2$. This cubic action takes the following form:
\begin{align}\label{eq:ac}
I&=\,\frac{i}{g^2} STr_{\mathfrak{osp}(1|32,\mathbb{R})} Tr_{\mathfrak{u}(N)} (M[M,M]_{\mathfrak{u}(N)})\,=\,
-\frac{1}{g^2}f^{abc}STr_{\mathfrak{osp}(1|32,\mathbb{R})}(M^a M^b M^c)\,=
\\
&=\,-\frac{1}{g^2}f^{abc}\left(Tr_{\mathfrak{sp}(32,\mathbb{R})}(m^a m^b m^c)+3i \Psi^{a\top} C m^b \Psi^c\right)\notag
\end{align}
which we can now express in terms of 12-dimensional representations,
where the symplectic matrix $m$ is given by (\ref{m12}).

Let us give a short overview of the steps involved in the computation of (\ref{eq:ac}). 
It amounts to performing traces of triple products of  $m^a$'s over $\mathfrak{sp}(32,\mathbb{R})$, i.e.
traces of products of Dirac matrices. We proceed by decomposing such products into their irreps using~(\ref{GamProd}).
The only contributions surviving the trace are those proportional to the unit matrix. 
Thus, the only terms left in (\ref{eq:ac}) will be those containing traces over triple products of $2$-forms, 
over products of a $2$-form and two $6$-forms, and over  triple products of $6$-forms, while
terms proportional to products of two $2$-forms and a $6$-form will yield zero contributions.

The two terms involving $Z^+$'s (to wit $CZ^+Z^+$ and $Z^+Z^+Z^+$) require some care, since 
$\Gamma^{A_1\ldots A_{12}}$ is proportional to $\Gamma_{\star}$ in $12D$, and hence 
$Tr({\cal P}^+\,\Gamma^{A_1\ldots A_{12}})\propto Tr(\Gamma^2_{\star})\neq 0$. 
Since double products of six-indices Gamma matrices decompose into $\Id$ and Gamma matrices with
2, 4 up to 12 indices, their trace with $\Gamma^{MN}$ will keep terms with 2, 10
or 12 indices (the last two containing Levi-Civita tensors) while their trace with $\Gamma^{M_1 \dots M_6}$   
will only keep those terms with 6, 8, 10 and 12 indices.

Finally, putting everything together, exploiting the self-duality of $Z^+$ 
and rewriting cubic products of fields contracted by $f^{abc}$ as a trace over 
$\mathfrak{u}(N)$, we get:
\begin{align}
I &=\, \frac{32i}{g^2}\,Tr_{\mathfrak{u}(N)}
\Bigg(
C_M^{\phantom{A}N}\, [C_N^{\phantom{A}O},C_O^{\phantom{A}M}]_{\mathfrak{u}(N)}\,-\,
\frac{1}{20}\, C_A^{\phantom{A}B}\, [Z^{+M_1\dots M_5}_{B}, 
Z_{M_1\dots M_5}^{+\phantom{\dots M_5}A}]_{\mathfrak{u}(N)}
\,+\notag\\
&+\,\frac{61}{2(3!)^3}\,Z^{+\phantom{BC}DEF}_{ABC}\,[Z^{+\phantom{EF}GHI}_{DEF}
,Z^{+\phantom{HI}ABC}_{GHI}]_{\mathfrak{u}(N)}\,+\notag \\
&+\,\frac{3i}{64}\Psi^{\top} C {\cal P}_+ \Gamma^{MN} [C_{MN}, \Psi]_{\mathfrak{u}(N)} \,+\,
\frac{3i}{32 \cdot 6!}\Psi^{\top} C {\cal P}_+ \Gamma^{M_1 \ldots M_6}[ Z^{+}_{M_1 \ldots M_6}, 
\Psi]_{\mathfrak{u}(N)}
\Bigg)\notag
\end{align}
where we have chosen: $\varepsilon^{0 \dots 11} = \varepsilon_{0 \dots 11} =+1$, since the metric contains
two time-like indices. Similarly, one can decompose invariant terms such as 
$STr_{\mathfrak{osp}(1|32,\mathbb{R})} Tr_{\mathfrak{u}(N)} (M^2)$ and \linebreak
$STr_{\mathfrak{osp}(1|32,\mathbb{R})} Tr_{\mathfrak{u}(N)} ([M,M]_{\mathfrak{u}(N)}[M,M]_{\mathfrak{u}(N)})$,
etc. While it might be interesting to investigate further the 12$D$ physics obtained
from such models and compare it to F-theory dynamics, we will not do so here. 
We will instead move to a detailed study of the better known $11D$ case, 
possibly relevant for M-theory.

\section{Study of the 11D M-theory case}

We now want to study the $11D$ matrix model more thoroughly. Similarly to the 12 dimensional case,
we embed the $SO(10,1)$ Clifford algebra into $\mathfrak{sp}(32,\mathbb{R})$ and replace the fundamental 
representation of $\mathfrak{sp}(32,\mathbb{R})$ by $SO(10,1)$ Majorana spinors.
A convenient choice of $32 \times 32$ Gamma matrices are the $\tilde{\Gamma}$'s we used in the 12$D$ case. 
We choose them as follows:
\begin{equation}
\tilde{\Gamma}^0=\begin{pmatrix}
0 & -\Id_{16} \\
\Id_{16} & 0 \\
\end{pmatrix}  \text{  ,  }
\tilde{\Gamma}^{10}=\begin{pmatrix}
0 & \Id_{16} \\
\Id_{16} & 0 \\
\end{pmatrix} \text{  ,  }
\tilde{\Gamma}^i=\begin{pmatrix}
\gamma^i & 0 \\
0 & -\gamma^i \\
\end{pmatrix}\phantom{a} \forall i = 1, \ldots, 9,
\end{equation}
where the $\gamma^i$'s build a Majorana representation of the
Clifford algebra of $SO(9)$, $\{\gamma^i,\gamma^j\} = 2
\delta^{ij} \Id_{16}$.
As before, we have $\tilde{\Gamma}^{10}= \tilde{\Gamma}^0 \tilde{\Gamma}^1
\ldots \tilde{\Gamma}^9$ provided $\gamma^1 \ldots \gamma^9 = \Id_{16}$, since
we can define $\gamma^9$ to be $\gamma^9 = \gamma^1 \ldots \gamma^8$.
This choice has $(\tilde{\Gamma}^0)^2=-\Id_{32}$, while $(\tilde{\Gamma}^M)^2=\Id_{32}$, 
$\forall M=1 \dots 10$ and gives a representation of $\{\tilde{\Gamma}^M,\tilde{\Gamma}^N\} = 2 \eta^{MN} \Id_{32}$ 
for the choice $(-,+,\ldots,+)$ of the metric. 
As we have again chosen all $\tilde{\Gamma}$'s to be real, this allows
to take $\tilde{B}=\Id$ in $\Psi^* = \tilde{B} \Psi$, which implies 
that the charge conjugation matrix is $\tilde{C}= \tilde{\Gamma}^{0}$.
Moreover, we have the following transposition rules for the $\tilde{\Gamma}$ matrices:
\begin{equation}\label{CGamn11}
\tilde{C} \tilde{\Gamma}^{M_1 \ldots M_n} \tilde{C}^{-1} = (-1)^{n(n+1)/2} 
(\tilde{\Gamma}^{M_1 \ldots M_n})^{\top}
\end{equation}
We will identify the fundamental representation of
$\mathfrak{sp}(32,\mathbb{R})$ with a 32-component Majorana
spinor of $SO(10,1)$. Splitting the 32 real components of the $\Psi$ into
$16+16$ as in:
$\Psi^{\top} = (\Phi_1^{\top},\Phi_2^{\top})$, we can use the following identity:
$$
(-i\Phi_2^{\top}, i\Phi_1^{\top}) = -i \Psi^{\top} \tilde{\Gamma}^0 = -i \Psi^{\top} \tilde{C} = -i \overline{\Psi}
$$
to write orthosymplectic matrices again as in~(\ref{mpsi}).

\subsection{Embedding of the 11D Super-Poincar\'e algebra in $\mathfrak{osp}(1|32,\mathbb{R})$ }
In 11$D$, we can also express the $\mathfrak{sp}(32,\mathbb{R})$ transformations in terms of translations, 
Lorentz transformations and new 5-form symmetries, by defining:
\begin{equation}
h = h_M P^M + h_{MN} J^{MN} + h_{M_1\ldots M_5} J^{M_1\ldots M_5}\;.
\end{equation}
With the help of~(\ref{GamComm}), we can compute this enhanced Super-Poincar\'e algebra 
as in dimension 12, using the following explicit representation of the generators:
\begin{equation}
P^M = \tilde{\Gamma}^{M} \,,\quad
J^{MN} =\frac{1}{2} \tilde{\Gamma}^{MN} \,,\quad
J^{M_1\ldots M_5} = \frac{1}{5!}\tilde{\Gamma}^{M_1\ldots M_5}
\end{equation}
In order to express everything in terms of the above generators, we need to dualize forms using the formula:
$\frac{1}{(11-k)!} \varepsilon^{M_1 \ldots M_{11}} \tilde{\Gamma}_{M_{k+1} \ldots 
M_{11}} = - \tilde{\Gamma}^{M_1 \ldots M_k}$.
This leads to the following superalgebra:
\begin{align}\label{AdS11}
[P^M,P^N] &\,=\, 4 J^{MN} \notag\\
[P^M,J^{OP}] &\,=\, 2 \eta^{M [O} P^{P]} \notag\\
[J^{MN},J^{OP}] &\,=\, -4 \eta^{[M [O} J^{N] P]} \notag\\
[P^M,J^{M_1\ldots M_5}] &\,=\, -\frac{2}{5!} \varepsilon^{M M_1\ldots
M_5}_{\phantom{M M_1\ldots M_5} N_1 \ldots N_5} J^{N_1\ldots N_5} \notag\\
[J^{MN},J^{M_1\ldots M_5}] &\,=\,-10\,\eta^{[M[M_1}\,J^{N]M_2\ldots M_5]} \notag\\ 
[J^{M_1\ldots M_5},J^{N_1\ldots N_5}] 
&\,=\,-\frac{2}{(5!)^2}\varepsilon^{M_1\ldots M_5 N_1\ldots
N_5}_{\phantom{M_1\ldots M_5 N_1\ldots N_5} A} P^A +\,\frac{1}{(3!)^2}\eta^{[M_1[N_1}\,\eta^{M_2\,N_2}\varepsilon^{M_3
\ldots M_5] N_3\ldots N_5]}_{\phantom{M_1\ldots M_5] N_1\ldots N_5]} O_1 
\ldots O_5} J^{O_1\ldots O_5} +\notag\\
&\,+\,\frac{1}{3!}\eta^{[M_1[N_1}\,\eta^{M_2\,N_2}\,\eta^{M_3\,N_3}\,\eta^{M_4\,N_4}\,J^{M_5]
N_5]}\\
[P^M,Q_A]\,&=\,(\tilde{\G}^M)^B_{\phantom{B}A} Q_B\notag\\
[J^{MN},Q_A]\,&=\, \frac{1}{2}(\tilde{\G}^{MN})^B_{\phantom{B}A} Q_B\notag\\
[J^{M_1\ldots M_5},Q_A]\,&=\, \frac{1}{5!}(\tilde{\G}^{M_1\ldots M_5})^B_{\phantom{B}A} Q_B
\notag\\
\{Q_A,Q^B\} \,&=\,\frac{i}{16} (\tilde{C} \tilde{\Gamma}_{M})_A^{\phantom{A}B} P^M -
\frac{i}{16} (\tilde{C} \tilde{\Gamma}_{MN})_A^{\phantom{A}B} J^{MN}
+ \frac{i}{16} (\tilde{C} \tilde{\Gamma}_{M_1\ldots M_5})_A^{\phantom{A}B} J^{M_1\ldots M_5}\,.\notag
\end{align}
Note that this algebra is the dimensional reduction from 12$D$ to 11$D$ of~(\ref{algebra}).
In particular, the first three lines build the $\mathfrak{so}(10,2)$ Lie algebra, but appear
in this new 11-dimensional context as the Lie algebra of symmetries of $AdS_{11}$ space 
(it is of course also the conformal algebra in 9+1 dimensions). We may wonder whether this 
superalgebra is a minimal supersymmetric extension of the $AdS_{11}$ Lie algebra or not. 
If we try to construct an algebra without the five-form symmetry generators,
the graded Jacobi identity forbids the appearence of a five-form central charge on the RHS of 
the $\{Q_A,Q^B\}$ anti-commutator. The number of independent components in this last line of the 
superalgebra will thus be bigger on the LHS than on the RHS. This is not strictly forbidden, but 
it has implications on the representation theory of the superalgebra. The absence of central 
charges will for example forbid the existence of shortened representations with a non-minimal
eigenvalue of the quadratic Casimir operator $C=-1/4 P_M P^M + J_{MN} J^{MN}$ (``spin'') of the 
$AdS_{11}$ symmetry group (see~\cite{Nic}). 
More generally, in 11D, either all objects in the RHS of the last line are central charges 
(this case corresponds simply to the $11D$ Super-Poincar\'e algebra) or they should all be 
symmetry generators.
Thus, although it is not strictly-speaking the minimal supersymmetric extension of the 
$AdS_{11}$ Lie algebra, it is certainly the most natural one. That's why some authors~\cite{BVP} 
call $\mathfrak{osp}(1|32,\mathbb{R})$ the super-$AdS$ algebra in $11D$. Here, we will 
stick to the more neutral $\mathfrak{osp}(1|32,\mathbb{R})$ terminology. 
Furthermore, $\mathfrak{osp}(1|32,\mathbb{R})$ is also the maximal ${\cal N}=1$ extension of the $AdS_{11}$
algebra. In principle, one could consider even bigger superalgebras, but we will not investigate
them in this article.

It is also worth remarking that similar algebras have been studied in~\cite{Sez} where they are called
topological extensions of the supersymmetry algebras for supermembranes and super-5-branes. 

\subsection{The supersymmetry properties of the 11D matrix fields}
Let us now look at the action of supersymmetries on the fields of an 
$\mathfrak{osp}(1|32,\mathbb{R})$ eleven-dimensional matrix model.
We expand once again the bosonic part of our former matrix
$M$ on the irrep of $SO(10,1)$ in terms of 32-dimensional $\Gamma$ matrices:
$$
m\,=\,
X_M\,\widetilde{\Gamma}^M\,+\,\frac{1}{2!}\,C_{MN}\,\widetilde{\Gamma}^{MN}
\,+\,\frac{1}{5!}\,Z_{M_1\ldots M_5}\,\widetilde{\Gamma}^{M_1\ldots M_5}\;,
$$ 
where the vector, the 2- and 5-form are given by:
$$
X_M\,=\, \frac{1}{32}\,Tr_{\mathfrak{sp}(32,\mathbb{R})}(m\,\widetilde{\Gamma}_{M})\,, \;
C_{MN}\,=\, -\frac{1}{32}\,Tr_{\mathfrak{sp}(32,\mathbb{R})}(m\,\widetilde{\Gamma}_{MN})\,, \;
Z_{M_1\ldots M_5}\,=\, \frac{1}{32}\,Tr_{\mathfrak{sp}(32,\mathbb{R})}
(m\,\widetilde{\Gamma}_{M_1\ldots M_5})\;.
$$

Let us give the whole $\delta^{(1)}_H$ transformation acting
on the fields (using the cyclic property of the trace, for instance: 
$Tr([h,m]\widetilde{\Gamma}^{M})= Tr(h[m,\widetilde{\Gamma}^{M}])$):
\begin{align*}
\delta^{(1)}_H\,X^M
&\,=\, 
2\left( h^{MQ}\,X_Q\,+\,h^{Q}\,C_Q^{\phantom{N}M}\,-\,
\frac{1}{(5!)^2}\,\varepsilon^{M M_1\ldots M_5}_{\phantom{M M_1\ldots M_5}N_1\ldots N_5}h^{N_1\ldots N_5}\,
Z_{M_1\ldots M_5} \right)
\,-\,\frac{i}{16}\,\chi^{\top}\widetilde{\Gamma}^0
\widetilde{\Gamma}^{M}\Psi
\;,\\
\delta^{(1)}_H\,C^{MN}
&\,=\, 
-4\,\left(h^{[M} X^{N]}\,-\,h^{[M}_{\phantom{[M}Q}C^{N]Q}
\,+\, \frac{1}{4!}\,h_{M_1\ldots M_4}^{\phantom{M_1\ldots M_4}[M}\,Z^{N] M_1\ldots M_4}
\right)\,+\,\frac{i}{16}\,\chi^{\top}\widetilde{\Gamma}^0 \widetilde{\Gamma}^{MN}\Psi\;,\\
\delta^{(1)}_H\,Z^{M_1\ldots M_5}
&\,=\,2 \left( \frac{1}{5!}
\varepsilon_{\phantom{M_1\ldots M_5}N_1\ldots N_5 Q}^{M_1\ldots M_5}\,h^{N_1\ldots N_5}\,X^{Q}
\,+\,5\,h_Q^{\phantom{Q}[M_1\ldots M_4}\,
C^{M_5]Q}\,-\,
5\,h_Q^{\phantom{Q}[M_1}\,Z^{M_2\ldots M_5]Q}\right.+\\
&\left.+\,\frac{1}{5!}\,\varepsilon^{M_1\ldots M_5}_{\phantom{M_1\ldots M_5}O 
N_1\ldots N_5}\,h^O\,Z^{N_1\ldots N_5}-\, \frac{1}{3\cdot4!} h^{O_1 \ldots O_5}\, \varepsilon_{O_1 \ldots O_5 N_1 N_2 N_3}^{\phantom{O_1\ldots O_5 N_1 N_2 N_3}[M_1 M_2 M_3}
\,Z^{M_4 M_5]N_1 N_2 N_3}\right)\,-\\
&\,-\,\frac{i}{16}\,\chi^{\top}\widetilde{\Gamma}^0
\widetilde{\Gamma}^{M_1 \ldots M_5}\Psi\;, \\
\delta^{(1)}_H\,\Psi &\,=\, \Big(h_M\,\widetilde{\Gamma}^M \,+\,h_{MN}\,\widetilde{\Gamma}^{MN}
\,+\,h_{M_1\ldots M_5}\,\widetilde{\Gamma}^{M_1\ldots M_5}\Big) \Psi\,-\notag\\
&\,-\,\widetilde{\Gamma}^{M} \chi X_M
\,-\,\frac{1}{2}\widetilde{\Gamma}^{MN} \chi C_{MN} 
\,-\,\frac{1}{5!}\widetilde{\Gamma}^{M_1\ldots
M_5} \chi Z_{M_1 \ldots M_5}\:,
\end{align*}
where the part between parentheses describes the symplectic transformations, while the rest represents
the supersymmetry variations.
Note that we used $\frac{1}{(11-k)!} \varepsilon^{M_1 \ldots M_{11}} \tilde{\Gamma}_{M_{k+1} \ldots 
M_{11}} = - \tilde{\Gamma}^{M_1 \ldots M_k}$ in $\delta^{(1)}_H\,Z^{M_1\ldots M_5}$ to dualize the 
Gamma matrices when needed.

\subsection{11-dimensional action for a supersymmetric matrix model} 
As in the $12D$ case, we will now consider a specific model, invariant
under $U(N)$ gauge and $\mathfrak{osp}(1|32,\mathbb{R})$ transformations.
The simplest such model containing interactions and ``propagators'' is a
cubic action along with a quadratic term. 
Hence, we choose:
\begin{align}\label{action11a}
I & =\, STr_{\mathfrak{osp}(1|32,\mathbb{R})\otimes\mathfrak{u}(N)} 
\left(-\,\mu\, M^2 + \frac{i}{g^2}\,M[M,M]_{\mathfrak{u}(N)}\,\right)\;.
\end{align}
Contrary to a purely cubic model, one loses invariance under $M \rightarrow M+K$ for a constant
diagonal matrix $K$, which contains the space-time translations of the BFSS model.
In contrast with the BFSS theory, our model doesn't exhibit the symmetries of flat $11D$ Minkowski 
space-time, so we don't really expect this sort of invariance. 
However, the symmetries generated by $P^M$ remain unbroken, as well as all other 
$\mathfrak{osp}(1|32,\mathbb{R})$ transformations.  Indeed, the related
bosonic part of the algebra (\ref{AdS11}) contains the symmetries of $AdS_{11}$
as a subalgebra, and as was pointed out in~\cite{Gao} and~\cite{Cham}, massive
matrix models with a tachyonic mass-term for the coordinate $X$'s fields appear 
in attempts to describe gravity in de Sitter spaces (an alternative approach can
be found in~\cite{Li}). Note that we take the opposite sign for
the quadratic term of (\ref{action11a}), this choice being motivated by the belief that $AdS$ vacua are 
more stable than $dS$ ones, so that the potential energy for physical bosonic 
fields should be positive definite in our setting.   

The computation of the 11-dimensional action for this supermatrix model is analogous 
to the one performed in 12 dimensions. We remind the reader that each entry of the matrix $M$ 
now becomes a hermitian matrix in the Lie algebra of $\mathfrak{u}(N)$ for some
large value of $N$ whose generators are defined as in the $12D$ case.

After performing in (\ref{action11a}) the traces on products of
Gamma matrices, it comes out that the terms of the form $XXX$,
$XXZ$, $XCC$, $CCZ$ and $XCZ$ have vanishing trace (since products of Gamma matrices related to these terms 
have decomposition in irreducible tensors that do not contain a term proportional to $\Id_{32}$) 
so that only terms of the form $XXC$, $XZZ$, $CZZ$, $CCC$, $ZZZ$ will remain
from the cubic bosonic terms. As for  terms containing fermions and the mass terms,
they are trivial to compute. Using~(\ref{GamProd}) and the usual duality relation for Gamma matrices in $11D$, 
one finally obtains the following result:
\begin{align}\label{axxion}
I&=\,-\, 32 \mu \,Tr_{\mathfrak{u}(N)}\left\{ X_M X^M\,-\,\frac{1}{2!}\,C_{MN}C^{MN}\,+\,\frac{1}{5!}\,Z_{M_1\ldots M_5}
Z^{M_1\ldots M_5}\,+\,\frac{i}{16}\overline{\Psi}\Psi\right\}\,+\notag\\
&\,+\frac{32i}{g^2}\,Tr_{\mathfrak{u}(N)}\Bigg(3\,C_{NM}\,[X^{M},X^{N}]_{\mathfrak{u}(N)}\,-\,
\varepsilon^{M_1\ldots M_{11}}
\left\{\frac{3}{(5!)^2}\,Z_{M_1\ldots M_5}\,[X_{M_{6}},Z_{M_7\ldots M_{11}}]_{\mathfrak{u}(N)}
\,-\, \right.\notag\\ 
&\,-\ \left.\frac{2^3\,5^2}{(5!)^3}\,Z_{M_1\,M_2\,M_3}^{\phantom{M_1\,M_2\,M_3}AB}\,[
Z_{AB\,M_4\,M_5\,M_6},Z_{M_7\ldots M_{11}}]_{\mathfrak{u}(N)}\right\}
\,+\,\frac{3}{4!}\,C_{MN}\,[Z_{A_1\ldots A_4}^{\phantom{A_1\ldots A_4}N},Z^{A_1\ldots A_4\,M}]_{\mathfrak{u}(N)}\,+\,\notag\\
&\,+\,C_{MN}\,[C^{N}_{\phantom{N}O},C^{OM}]_{\mathfrak{u}(N)}
\,+\,\frac{3i}{32}\left\{\overline{\Psi}\,\widetilde{\Gamma}^M[X_M,\Psi]_{\mathfrak{u}(N)} \,+\,
\frac{1}{2!}\,\overline{\Psi}\,\widetilde{\Gamma}^{MN}\,[C_{MN},\Psi]_{\mathfrak{u}(N)}\,+\,\right.\\
&\,+\,\left. \frac{1}{5!}\, \overline{\Psi}\,\widetilde{\Gamma}^{M_1\ldots M_5}\,[
Z_{M_1\ldots M_5},\Psi]_{\mathfrak{u}(N)} \right\}\Bigg)\;.\notag
\end{align}

\section{Dynamics of the $11D$ supermatrix model and its relation to BFFS theory}

Now, we will try to see to what extent our model may describe at least part
of the dynamics of M-theory. Since the physics of the BFSS matrix model and its relationships 
to 11D supergravity and superstring theory are relatively well understood, if our model is 
to be relevant to M-theory, we expect it to be related to BFSS theory at least in some regime. 
To see such a relationship, we should reduce our model to one of its ten-dimensional sectors
and turn it into a matrix quantum mechanics.

\subsection{Compactification and T-duality of the $11D$ supermatrix action}

If we want to link (\ref{axxion}) to BFSS, which is basically a quantum
mechanical supersymmetric matrix model, we should reduce the eleven-dimensional target-space 
spanned by the $X^M$'s to 10 dimensions, and, at the same time, let a ``time'' parameter $t$ appear. 
At this stage, the world-volume of the theory is reduced to one point. We start by decompactifying it 
along two directions, following the standard procedure outlined in \cite{Wati}. Namely, we compactify
the target-space coordinates $X_0$ and $X_{10}$ on circles of respective radii $R_0=R$ and $R_{10}=\omega R$.
We introduce the rescaled field $X'_{10} \equiv X_{10}/\omega$ which has the same $2\pi R$ periodicity as $X_0$. 
We can then perform T-dualities on $X_0$ and $X'_{10}$ to circles of dual radii $ \widehat{R} \equiv l^2_{11}/R$ 
(parametrized by $\tau$ and $y$), where $l_{11}$ is some scale, typically the 11-dimensional Planck length. 
The fields of our theory, for simplicity denoted here by $Y$, now depend on the world-sheet coordinates $\tau$
and $y$ as follows: 
\begin{equation}  \label{eq:compk}
Y(\tau,y)\;=\; \sum_{m,n} Y_{mn}\,e^{i (m\tau+ny)/\widehat{R}} \;.
\end{equation}
As a consequence, we now need to average the action over $\tau$ and $y$
with the measure $d\tau dy/(2\pi\widehat{R})^2$. Finally, one should identify under T-duality:
\begin{equation}  \label{eq:compk2}
X_0\sim 2\pi l^2_{11}\,\Big(i\partial_{\tau}-A_{\tau}(\tau,y)\Big)\triangleq i\mathcal{\widehat{D}}_{\tau}\;,\quad
X_{10} \equiv \omega X'_{10} \sim 2\pi\omega l^2_{11}\,\Big(i\partial_{y}-A_{y}(\tau,y)\Big)\triangleq i \omega 
\mathcal{\widehat{D}}_y\;,
\end{equation}
where $A_{\tau}$ and $A_y$ are the connections on the $U(N)$ gauge bundle over the world-sheet. 
For notational convenience, we rewrite $\phi \,\triangleq \,C_{0\,10}$, 
$F_{\tau y} \, \triangleq \, -i \, [\mathcal{\widehat{D}}_{\tau},\mathcal{\widehat{D}}_{y}]$ and 
$\widetilde{\Gamma}_{*}\triangleq\widetilde{\Gamma}_{10} $ and encode the possible values of the indices 
in the following notation: 
\begin{align*}
A,\; B\,=\,0,\ldots ,10\;, &\qquad i,\;j,\;k\,=\,1,\ldots ,9\;, \\
\alpha\,=\,1,\ldots ,10\;, &\qquad \beta\,=\,0,\ldots ,9\;.
\end{align*}
Then, the compactified version of (\ref{axxion}) reads:
\begin{align} \label{eq:comp}
I_c &= \frac{32i}{g^2}\,\int\,\frac{d\tau dy}{(2\pi \widehat{R})^2} 
\,Tr_{\frak{u}(N)}\Bigg( -6\,C_{i0}\,i[\mathcal{\widehat{D}}_{\tau},X_i]\,
+\,6\omega\, C_{i10}\,i[\mathcal{\widehat{D}}_{y},X_i]\,
+\, \frac{3}{32}\,\overline{\Psi}\,\widetilde{\Gamma}_{0}\,[\mathcal{\widehat{D}}_{\tau},\Psi]\,-\nonumber\\ 
&-\, \frac{3 \omega}{32}\,\overline{\Psi}\,\widetilde{\Gamma}_{*}\,[\mathcal{\widehat{D}}_y,\Psi]\,
-\,\frac{3}{(5!)^2}\,\varepsilon_{\alpha_1\cdots \alpha_{10}0}\,
Z_{\alpha_1\cdots \alpha_5}\, i[\mathcal{\widehat{D}}_{\tau},Z_{\alpha_6\cdots \alpha_{10}}]\,
+\,\frac{3\omega}{(5!)^2}\,\varepsilon^{\beta_1\cdots \beta_{10}10}\,
Z_{\beta_1\cdots \beta_5}\, i[\mathcal{\widehat{D}}_{y},Z_{\beta_6 \cdots \beta_{10}}]\,+\nonumber\\
&+\,6i\omega\, \phi \, F_{\tau y}\, +\,3\,C_{ij}\,[X_j,X_i]\,
+\, \frac{3}{(5!)^2}\,\varepsilon^{A_1\cdots A_{10}}_{\phantom{A_1\cdots A_{10}}j}\, 
Z_{A_1\cdots A_5}\,[X_j,Z_{A_6\cdots A_{10}}]\,-\nonumber \\
&-\,\frac{2^3 5^2}{(5!)^3}\,\varepsilon^{A_1\cdots A_{11}}\, Z_{A_1 A_2
A_3}^{\phantom{M_1 M_2 M_3}B_1 B_2}\,[Z_{B_1 B_2 A_4 A_5 A_6}, Z_{A_7\cdots
A_{11}}] \,+\, \frac{3}{4!}\left\{ C_{ij}\,[Z_{j\, A_1 \cdots A_4}, Z_{i}^{
\phantom{i} A_1\cdots A_4}]\,-\, \right.  \nonumber \\
&-\, \left. 2\,C_{i0}\,[Z_{0 \,\alpha_1\cdots \alpha_4}, Z_{i\,\alpha_1
\cdots \alpha_4}]\,+\, 2\,C_{i10}\,[Z_{10\, \beta_1\cdots \beta_{4}}, Z_{i}^{
\phantom{i} \beta_1\cdots \beta_{4}}] \,-\,2\,\phi\,[Z_{10\, i_1\cdots
i_{4}},Z_{0\, i_1\cdots i_{4}}]\right\}\,+\,  \nonumber \\
&+\,C_{ij}\,[C_{jk},C_{ki}]\,+\,3\,C_{i0}\,[C_{k0},C_{ki}]\,-\,3\,C_{i10}
\,[C_{k 10},C_{ki}] \,+\, 6\,\phi\,[C_{k10},C_{k0}]\,+\,  \\
&+\,\frac{3i}{32}\,\Big\{ \overline{\Psi}\,\widetilde{\Gamma}
_{i}\,[X_{i},\Psi] \,+\, \frac{1}{2!}\,\overline{\Psi}\,
\widetilde{\Gamma}_{ij}[C_{ij},\Psi]\,-\, \overline{\Psi}\,
\widetilde{\Gamma}_{i}\widetilde{\Gamma}_{0}\,[C_{i0},\Psi]\,+\, \overline{\Psi}\,
\widetilde{\Gamma}_i\widetilde{\Gamma}_{*}\,[C_{i10},\Psi]\,
-\,\overline{\Psi}\,\widetilde{\Gamma}_{0}\widetilde{\Gamma}_{*}\,[\phi,\Psi]\,+\nonumber\\
&+\,\frac{1}{5!}\, \overline{\Psi}\,\widetilde{\Gamma}^{A_1\cdots
A_5}\,[Z_{A_1\cdots A_5},\Psi]\Big\}\,+
\,i\mu g^2 \Big(\mathcal{\widehat{D}}_{\tau}\mathcal{\widehat{D}}_{\tau}\,
-\, \omega^2 \,\mathcal{\widehat{D}}_y\mathcal{\widehat{D}}_y\,
+\,X_{i}X_{i}\,+\,\frac{i}{16}\overline{\Psi }\Psi \,+\,\phi ^{2}\,-\nonumber\\
&-\,\frac{1}{2!}\,C_{ij}C_{ij}\,+\,C_{i0}C_{i0}\,-
\,C_{i10}C_{i10}\,+\,\frac{1}{5!}Z_{A_{1}\cdots A_{5}}Z^{A_{1}\cdots A_{5}}
\Big)
\Bigg)\;.\notag
\end{align}

Repeated indices are contracted, and when they appear alternately up and
down, Minkowskian signature applies, whereas Euclidian signature is in force
when both are down.

\subsection{Ten-dimensional limits and IMF}
Since the BFSS matrix model is conjectured to describe M-theory in the infinite momentum frame,
we shall investigate our model in this particular limit. For this purpose, let's define 
the light-cone coordinates $t_+ \equiv (\tau + y)/\sqrt{2}$ and $t_- \equiv (\tau - y)/\sqrt{2}$ 
and perform a boost in the $y$ direction. In the limit where the boost parameter $u$ is large,
the boost acts as $(t_{+},t_{-}) \xrightarrow{\sim} (u t_{+}, u^{-1} t_{-})$, or as
$(\tau,y) \xrightarrow{\sim} \sqrt{2} (u t_{+}, u t_{+})$ on the original coordinates.
In particular, when $u \rightarrow \infty$, the $t_{-}$ dependence disappears from the action
and we can perform the trivial $t_{-}$ integration. The dynamics is now solely described by the parameter 
$t \equiv \sqrt{2} u t_{+}$, which is decompactified through this procedure. In particular, both 
$\mathcal{\widehat{D}}_{\tau}$ and $\mathcal{\widehat{D}}_y$ are mapped into $\mathcal{\widehat{D}}_{t}$.

So far, the ratio of the compactification radii $\omega$ is left undetermined and it parametrizes a continuous
family of frames. It affects the kinetic terms as:
\begin{align} \label{eq:comp2}
I_c &= \frac{32i}{g^2}\,\lim_{u\rightarrow \infty} \int_{-\pi\widehat{R}u}^{\pi\widehat{R}u}\,
\frac{dt}{2 \sqrt{2} \pi \widehat{R}u} \,Tr_{\frak{u}(N)}\Bigg( -6\,\Big(C_{i0}-\omega\, C_{i10}\Big)\,
i[\mathcal{\widehat{D}}_t,X_i]\,+\, \frac{3}{32}\,\overline{\Psi}\,\Big(\widetilde{\Gamma}_{0}-
\omega\,\widetilde{\Gamma}_{*}\Big)\,[\mathcal{\widehat{D}}_t,\Psi]\,-\nonumber\\ 
-&\,\frac{3}{(5!)^2}\,\varepsilon_{\alpha_1\cdots \alpha_{10}0}\,
Z_{\alpha_1\cdots \alpha_5}\, i[\mathcal{\widehat{D}}_t,Z_{\alpha_6\cdots \alpha_{10}}]\,
+\,\frac{3\omega}{(5!)^2}\,\varepsilon^{\beta_1\cdots \beta_{10}10}\,
Z_{\beta_1\cdots \beta_5}\, i[\mathcal{\widehat{D}}_t,Z_{\beta_6 \cdots \beta_{10}}]\,+\, ... \,\Bigg)
\end{align} 
In order to have a non-trivial action, as in the BFSS case, we must take the limit $u \rightarrow \infty$
together with $N \rightarrow \infty$ in such a way that $N/(\widehat{R}u) \rightarrow \infty$.
In the following, we will write $\overline{R} \equiv \widehat{R}u$, implicitly take the limit 
$(\overline{R},N) \rightarrow \infty $ and let $t$ run from $-\infty$ to $\infty$.
 
In the usual IMF limit, one starts from an uncompactified $X_0$. In our notation, this corresponds to 
$R \rightarrow \infty$, i.e. to the particular choice $\omega=R_{10}/R \rightarrow 0$. So, in the IMF limit,
all terms proportional to $\omega$ drop out of (\ref{eq:comp2}). In the following chapters, we will restrict
ourselves to this case, since we are especially interested in the physics of our model in the infinite
momentum frame.

\subsection{Dualization of the mass term}

Let us comment on the meaning of the $\mathcal{\widehat{D}}_t^{2}$ term arising from the T-dualization of 
the mass term $Tr((X_0)^2)$, which naively breaks gauge invariance. To understand how it works, we should 
recall that the trace is defined by the following sum:
\begin{equation}\label{covconst}
Tr_{\frak{u}(N)}(-\mathcal{\widehat{D}}_t^{2})\,=\,-\sum_{a}\langle
u_{a}(t)|\mathcal{\widehat{D}}_{t}^{2}|u_{a}(t)\rangle \,=\,\sum_{a}\Vert
i\mathcal{\widehat{D}}_{t}|u_{a}(t)\rangle \Vert ^{2}\quad .  
\end{equation}
for a set of basis elements $\left\{|u_{a}(t)\rangle \right\}_{a}$ of $\frak{u}(N)$, 
which might have some $t$-dependence or not. If the $|u_{a}(t)\rangle$ are covariantly constant, 
the expression (\ref{covconst}) is obviously zero. Choosing the $|u_{a}(t)\rangle $ to be covariantly
constant seems to be the only coherent possibility. Such a covariantly constant basis is:
\[
|u_{a}(t)\rangle \,\triangleq \,e^{-i\int_{t_{0}}^{t}A_{0}(t^{\prime
})\,dt^{\prime }}|u_{a}\rangle \quad ,
\]
(where the $|u_{a}\rangle $'s form a constant basis, for instance, the generators of 
$\frak{u}(N)$ in the adjoint representation). Now, $t$ lives on a circle  and the function 
$\exp i\int_{t_{0}}^{t}A_{0} (t^{\prime})\,dt^{\prime}$ is well-defined only if the zero-mode 
$A_{0}^{(0)}=2\pi n$, $n\in \Z$. But we can always set $A_{0}^{(0)}$ to zero, since it doesn't 
affect the behaviour of the system, as it amounts to a mere constant shift in ''energy''. 
With this choice, we can integrate $\mathcal{\widehat{D}}_t$ by part without worrying about the trace.

\subsection{Decomposition of the 5-forms}

In (\ref{eq:compk2}), the only fields to be dynamical are the $X_{i}$, the 
$Z_{\alpha_{1}\cdots \alpha _{5}}$ and the $\Psi $. The remaining ones are either the conjugate 
\textit{momentum}-like fields when they multiply derivatives of dynamical fields, or 
\textit{constraint}-like when they only appear algebraically.

Thus, the $C_{i0}$ and $\overline{\Psi }$ have a straightforward
interpretation as \textit{momenta} conjugate respectively to the $X_{i}$ and
to $\Psi $. For the 5-form fields $Z_{A_{1}\cdots A_{5}}$ however, the
matter is a bit more subtle, due to the presence of the $11D$ $\varepsilon $~tensor 
in the kinetic term for the 5-form fields. Actually,
the real degrees of freedom contained in $Z_{A_{1}\cdots A_{5}}$ decompose as follows, 
when going down from $(10+1)$ to $9$ dimensions: 
\begin{eqnarray}\label{decombre}
\Omega ^{5}(\mathcal{M}_{10,1},\mathbb{R}) &\longrightarrow &
3\times\Omega ^{4}(\mathcal{M}_{9},\mathbb{R})\oplus \Omega ^{3}(\mathcal{M}_{9},\mathbb{R}
)\quad .
\end{eqnarray}
To be more specific (as in our previous convention, $i_{k}=1,\ldots , 9$ are
purely spacelike indices in $9D$), the 3-form fields on the RHS of~(\ref{decombre}) are 
$Z_{i_{1}i_{2}i_{3}0,10}\triangleq B_{i_{1}i_{2}i_{3}}$, while the 4-form fields are
$Z_{i_{1}i_{2}i_{3}i_{4}10}\triangleq Z_{i_{1}i_{2}i_{3}i_{4}}$, 
$Z_{i_{1}i_{2}i_{3}i_{4}0}\triangleq H_{i_{1}i_{2}i_{3}i_{4}}$ and\footnote{Using
$$
\varepsilon ^{j_{1}\cdots j_{N}i_{N+1}\cdots i_{9}0,10}\,\,\varepsilon
_{k_{1}\cdots k_{N}i_{N+1}\cdots i_{9}0,10}\,=\,-(9-N)!\sum_{\pi }\sigma
(\pi )\prod_{n=1}^{N}\delta _{k_{\pi (n)}}^{j_{n}}\quad ,  
$$ where $\pi $ is any permutation of $(1,2,..,N)$ and $\sigma (\pi )$ is the
signature thereof, this relation can be inverted:
$
Z_{i_{1}\cdots i_{5}}\,=\,\frac{1}{4!}\,\varepsilon _{i_{1}\cdots
i_{5}j_{6}\cdots j_{9}}\,\Pi ^{j_{6}\cdots j_{9}}\quad ,
$}
$\Pi^{i_{1}\cdots i_{4}}\,\triangleq \,1/5!\,\varepsilon ^{j_{1}\cdots
j_{5}i_{1}\cdots i_{4}0,10}Z_{j_{1}\cdots j_{5}};$ these conventions allow us
to cast the kinetic term for the 5-form fields into the expression
$6/4!\,\Pi^{i_{1}\cdots i_{4}}\,[\mathcal{\widehat{D}}_{t},Z_{i_{1}\cdots i_{4}}]$,
while $B$ and $H$ turn out to be \textit{constraint}-like
fields, the whole topic being summarized in Table~1.

\begin{center}
\begin{tabular}{c|c|c||c|c}
\emph{dynamical var.} & \emph{number of real comp.} & \emph{conjugate momenta} & \emph{
constraint-like} & \emph{number of real comp.} \\ \hline
$X_{i}$ & 9 & $C_{i0}$ & $C_{ij}$ & 36 \\ 
&  &  & $C_{i10}$ & 9 \\ 
&  &  & $\phi $ & 1 \\ \hline
$Z_{i_{1}\cdots i_{4}}$ & 126 & $\Pi _{i_{1}\cdots i_{4}}$ & $H_{i_{1}\cdots
i_{4}}$ & 126 \\ 
 & &  &  $B_{i_{1}i_{2}i_{3}}$ & 84 \\ 
\hline
$\displaystyle{\Psi}$ & 32  & $\displaystyle{\overline{\Psi }}$ &  &  \\ \hline
\end{tabular}\vspace{0.5cm}
Table 1: \textit{Momentum}-like and \textit{constraint}-like auxiliary fields
\end{center}
We see that longitudinal 5-brane degrees of freedom are described by the 4-form $Z_{i_{1}\cdots i_{4}}$, while 
transverse 5-brane fields $Z_{i_{1}\cdots i_{5}}$ appear in the definition of the conjugate momenta.  
As they are dual to one another, we could also have exchanged their respective r\^oles.
Both choices describe the same physics. We can thus interpret these degrees of freedom as transverse 5-branes, 
completing the BFSS theory, which already contains longitudinal 5-branes as bound states of D0-branes.  
                          
Choosing the $\varepsilon_{i_{1}\cdots i_{9}}$ tensor in 9 spatial dimensions to be: 
\[
\varepsilon _{i_{1}\cdots i_{9}}\,\triangleq \,\varepsilon _{i_{1}\cdots
i_{9}}^{\phantom{i_1\cdots i_9}0,10}\,=\,-\varepsilon _{i_{1}\cdots i_{9}0,10}\quad ,
\]
we can express the action $I_{c}$ in terms of the degrees of freedom appearing in Table~1 
(note that from now on all indices will be down, the signature for squared expressions is Euclidean
and we write $\mathcal{D}_{t}$ instead of $\widehat{\mathcal{D}}_{t}$ ):
\begin{align}\label{acfin}
I_{c}\,& =\,\frac{8\sqrt{2}i}{\pi g^{2}\overline{R}}\,\int \,dt\,Tr_{\frak{u}(N)}\Bigg( 
-6i\,C_{i0}\,[\mathcal{D}_{t},X_{i}]\,-\,
\frac{i}{4}\,\Pi_{i_{1}\cdots i_{4}}\,[\mathcal{D}_{t},Z_{i_{1}\cdots i_{4}}]\,+\,
\frac{3}{32}\,\overline{\Psi }\,\widetilde{\Gamma }_{0}\,[\mathcal{D}_{t},\Psi]\,+\,
3\,C_{ij}\,[X_{j},X_{i}]\, - \nonumber \\
& +\left( \,\Pi_{i_{1}i_{2}i_{3}\,j}\,[X_{j},B_{i_{1}i_{2}i_{3}}]\,-\,\frac{1}{4\cdot4!}
\,\varepsilon_{i_{1}\cdots i_{8}\,j}Z_{i_{1}\cdots
i_{4}}\,[X_{j},H_{i_{5}\cdots i_{8}}]\right) +\,\frac{1}{3!\cdot 4!}\,W(Z,\Pi,H,B)\,+  \nonumber \\
& +\,\frac{1}{2}\left\{ C_{ij}\,K_{ij}(Z,\Pi ,H,B)\,-
2\,C_{i0}\,\left(\frac{1}{4\cdot 4!}\varepsilon_{i\,j_{1}\cdots j_{4}k_{1}\cdots k_{4}}
[H_{j_{1}\cdots j_{4}},\Pi _{k_{1}\cdots k_{4}}]\,+\,
[Z_{i\,j_{1}j_{2}j_{3}},B_{j_{1}j_{2}j_{3}}]\right) \,+\right. \nonumber \\
& \,+\,\left. 2\,C_{i10}\,\left(\frac{1}{4\cdot 4!}\varepsilon_{i\,j_{1}\cdots j_{4}k_{1}\cdots k_{4}}
[Z_{j_{1}\cdots j_{4}},\Pi _{k_{1}\cdots k_{4}}]\,-\,
[H_{i\,j_{1}j_{2}j_{3}},B_{j_{1}j_{2}j_{3}}]\right)\,-\,
\frac{1}{2}\,\phi \,[Z_{i_{1}\cdots i_{4}},H_{i_{1}\cdots i_{4}}]\right\} \,+\,  \nonumber \\
& +\,C_{ij}\,[C_{jk},C_{ki}]\,+\,3\,C_{i0}\,[C_{k0},C_{ki}]\,-\,3\,C_{i10}\,[C_{k10},C_{ki}]\,
+\,6\,\phi \,[C_{k10},C_{k0}]+\,  \nonumber \\
& +\,\frac{3i}{32}\,\left\{ \overline{\Psi }\,\widetilde{\Gamma }_{i}\,[X_{i},\Psi ]\,+\,
\frac{1}{2!}\,\overline{\Psi }\,\widetilde{\Gamma }_{ij}[C_{ij},\Psi ]\,-\,
\overline{\Psi }\,\widetilde{\Gamma }_{i}\widetilde{\Gamma }_{0}\,[C_{i0},\Psi ]\,+\,
\overline{\Psi }\,\widetilde{\Gamma }_{i}\widetilde{\Gamma }_{*}\,[C_{i10},\Psi ]\,-\,\right. \\
& -\,\left. \overline{\Psi }\,\widetilde{\Gamma }_{0}\widetilde{\Gamma }_{*}\,[\phi ,\Psi ]
\,+\,\frac{1}{4!}\,\overline{\Psi }\,\widetilde{\Gamma}_{i_1\cdots i_4}\widetilde{\Gamma}_{*}
[Z_{i_1\cdots i_4},\Psi ]\,+\,
\frac{1}{4!}\,\overline{\Psi}\,\widetilde{\Gamma}_{i_1\cdots i_4}\widetilde{\Gamma}_{0}
\widetilde{\Gamma}_{*}[\Pi_{i_1\cdots i_4},\Psi ]\,+\,\right. \nonumber 
\\
& -\,\left.\frac{1}{4!}\,\overline{\Psi}\,\widetilde{\Gamma}_{i_1\cdots i_4}
\widetilde{\Gamma}_{0}[H_{i_1\cdots i_4},\Psi ]\,-
\,\frac{1}{3!}\,\overline{\Psi}\,\widetilde{\Gamma}_{i_1 i_2 i_3}\widetilde{\Gamma}_{0}
\widetilde{\Gamma}_{*}[B_{i_1 i_2 i_3},\Psi]\right\}\,+
\,\mu g^{2}i\left\{ (X_{i})^2\,+\,\frac{i}{16}\,\overline{\Psi }\Psi \,+\,\phi ^{2}\,- \right.\nonumber \\ 
&\left. -\,\frac{1}{2!}\,(C_{ij})^2\,+\,(C_{i0})^2\,-\,(C_{i10})^2\,+
\,\frac{1}{4!}\left(\left( Z_{i_{1}\cdots i_{4}}\right) ^{2}\,+
\,\left( \Pi _{i_{1}\cdots i_{4}}\right) ^{2}\,-
\,\left( H_{i_{1}\cdots i_{4}}\right)^{2}\,-
\,4\left( B_{i_{1}i_{2}i_{3}}\right)^{2}\right)\right\}\Bigg)\;.\notag
\end{align}
We have redefined the two following lengthy expressions in a compact
way to cut short: first the term coupling the various 5-form components to the $C_{ij}$:
\begin{equation*}
K_{ij}(Z,\Pi ,H,B)\, \triangleq 
[Z_{j\,k_{1}k_{2}k_{3}},Z_{i\,k_{1}k_{2}k_{3}}]\,+
\,[\Pi _{j\,k_{1}k_{2}k_{3}},\Pi_{i\,k_{1}k_{2}k_{3}}]\,-
\,3[B_{j\,k_{1}k_{2}},B_{i\,k_{1}k_{2}}]\,-
\,[H_{j\,k_{1}k_{2}k_{3}},H_{i\,k_{1}k_{2}k_{3}}]\,,
\end{equation*}
and second, the trilinear couplings amongst the 5-form components:
\begin{eqnarray*}
W(Z,\Pi ,H,B)\, &\triangleq &\,\varepsilon_{i_{1}\cdots i_{9}}\,\bigg\{
B_{i_{1}i_{2}\,j}\,\left(2\,[\Pi_{j\,i_{3}i_{4}i_{5}},\Pi_{i_{6}\cdots i_{9}}]\,
-[Z_{j\,i_{3}i_{4}i_{5}},Z_{i_{6}\cdots i_{9}}]\,
-[H_{j\,i_{3}i_{4}i_{5}},H_{i_{6}\cdots i_{9}}]\right)+\\
&&+\,\frac{2}{3}\,B_{i_{1}i_{2}i_{3}}\,\left(\,[B_{i_{4}i_{5}i_{6}},B_{i_{7}i_{8}i_{9}}]\,
+\,[Z_{i_{4}i_{5}i_{6}\,j},Z_{j\,i_{7}i_{8}i_{9}}]\,
-\,[H_{i_{4}i_{5}i_{6}\,j},H_{j\,i_{7}i_{8}i_{9}}]\,\right)\bigg\}\\
&&+\,(3!)^2\,\Pi_{i_{1}i_{2}j_{1}j_{2}}\,[Z_{j_{1}j_{2}k_{1}k_{2}},H_{k_{1}k_{2}i_{1}i_{2}}] \,\quad. 
\end{eqnarray*}

\subsection{Computation of the effective action}

We now intend to study the effective dynamics of the $X_i$ and $\Psi$ fields, in
order to compare it to the physics of D0-branes as it is described by the BFSS matrix model.
For this purpose, we start by integrating out the 2-form momentum-like and constraint-like 
fields, which will yield an action containing the BFSS matrix model as its leading term with,
in addition, an infinite series of couplings between the fields. Similarly, one would like to integrate out
the $Z$-type momenta and constraints $\Pi$, $H$ and $B$, to get an effective action for the 5-brane 
(described by $Z_{ijkl}$) coupled to the D0-branes. We will however not do so in the present paper, 
but leave this for further investigation.

To simplify our expressions, we set:\footnote{If we consider $X$ and hence $C$, $Z$ and $\Psi$ to have 
the engineering dimension of a length, then so has $\beta$, while $\gamma$ has dimension (length)$^{-4}$.} 
\[
\beta \,\triangleq \,\mu g^{2}\quad ,\qquad \qquad \gamma \,\triangleq \,\frac{
8\sqrt{2}}{\pi g^{2}\overline{R}}\quad ,
\]
and write (\ref{acfin}) as (after taking the trace over $\mathfrak{u}(N)$):
\begin{equation}
I_{c}\,=\,\gamma \int dt\,\left( \beta (\mathbf{C}
_{i}^{a})^{\intercal }(\mathcal{J}_{ij}^{ab}+\Delta _{ij}^{ab})\mathbf{C}
_{j}^{b}\,+\,\mathbf{C}_{i}^{a}\cdot \mathbf{F}_{i}^{a}\,+\,\mathcal{L}
_{C}\,+\,\mathcal{L}_{\phi }\,+\,\widehat{\mathcal{L}}\right) \quad .
\label{actpert}
\end{equation}
For convenience, we have resorted to a very compact notation, where:
\begin{eqnarray*}
\mathbf{C}_{i}^{a}\, \triangleq \,\left( 
\begin{array}{c}
C_{i0}^{a} \\ 
C_{i10}^{a}
\end{array}
\right) \; , \qquad
\mathcal{J}_{ij}^{ab}\, \triangleq \,\left( 
\begin{array}{cc}
-\delta ^{ab}\delta _{ij} & 0 \\ 
0 & \delta ^{ab}\delta _{ij}
\end{array}
\right) \; , \qquad
\Delta _{ij}^{ab}\, \triangleq \,\frac{3f^{abc}}{\beta }\left( 
\begin{array}{cc}
C_{ij}^{c} & \phi ^{c}\delta _{ij} \\ 
-\phi ^{c}\delta _{ij} & -C_{ij}^{c}
\end{array}
\right) \quad ,
\end{eqnarray*}
and where the components of the vector $\mathbf{F}_{i}^{a}\,=\,\left( 
\begin{array}{c}
F_{i}^{a} \\ 
G_{i}^{a}
\end{array}
\right) $, are given by the following expressions:  
\begin{eqnarray*}
F_{i} &\triangleq &6\,[{\cal{D}}_t,X_{i}]\,-\,
\frac{i}{4\cdot 4!}\varepsilon _{i\,j_{1}\cdots j_{4}k_{1}\cdots k_{4}}\,
[H_{j_{1}\cdots j_{4}},\Pi _{k_{1}\cdots k_{4}}]\,-\,
i\,[Z_{i\,j_{1}j_{2}j_{3}},B_{j_{1}j_{2}j_{3}}]\,-\,
\frac{3}{32}\{\overline{\Psi },\widetilde{\Gamma }_{i}\widetilde{\Gamma }_{0}\Psi\}\quad , \\
G_{i} &\triangleq & \frac{i}{4\cdot4!}\varepsilon_{i\,j_{1}\cdots j_{4}k_{1}\cdots k_{4}}
[Z_{j_{1}\cdots j_{4}},\Pi_{k_{1}\cdots k_{4}}]\,-\,i\,
[H_{i\,j_{1}j_{2}j_{3}},B_{j_{1}j_{2}j_{3}}]\,+\,
\frac{3}{32}\{\overline{\Psi },\widetilde{\Gamma }_{i}\widetilde{\Gamma }_{*}\Psi\} \quad .
\end{eqnarray*}
Note that we have written $if^{abc}\overline{\Psi }^b \widetilde{\Gamma }_{\ldots}\Psi^c$ as
$\{\overline{\Psi },\widetilde{\Gamma}_{\ldots}\Psi\}^a$ with a slight abuse of notation. 
The remaining terms in the action (\ref{actpert}) depending on $C_{ij}$ and $\phi $ are contained in
\begin{eqnarray*}
\mathcal{L}_{C} &\triangleq &\frac{\beta }{2}\,(C_{ij}^{a})^{2}\,+
\,E_{ij}^{a}C_{ij}^{a}\,-\,f^{abc}C_{ij}^{a}C_{jk}^{b}C_{ki}^{c}\quad , \\
\mathcal{L}_{\phi } &\triangleq &-\,\beta (\phi ^{a})^{2}\,+\,J^{a}\phi
^{a}\quad ,
\end{eqnarray*}
with the following definitions
\begin{eqnarray*}
E_{ij} &\triangleq &\,\frac{i}{2}K_{ij}\,+\,3i\,[X_{i},X_{j}]\,+\,
\frac{3}{64}\{\overline{\Psi },\widetilde{\Gamma }_{ij}\Psi\} \quad , \\
J &\triangleq &\,\frac{-i}{4}[Z_{i_{1}\cdots i_{4}},H_{i_{1}\cdots i_{4}}]\,-\,
\frac{3}{32}\{\overline{\Psi },\widetilde{\Gamma }_{0}\widetilde{\Gamma }_{*}\Psi\} \quad ,
\end{eqnarray*}
and finally $\,\widehat{\mathcal{L}}$ is the part of $I_{c}$ in (\ref{acfin}) independent of $C_{ij}$,
$C_{i10}$, $C_{i0}$ and $\phi$. 
in other words the part containing only dynamical fields (fermions $\Psi $ and coordinates $X_{i}$) 
as well as all fields related to the 5-brane (the dynamical ones: $Z$ and $\Pi $, as well
as the constrained ones: $B$ and $H$).

Now, (\ref{actpert}) is obviously bilinear in the $\mathbf{C}_{i}^{a}$ (note
that $\Delta _{ij}^{ab}$ \textit{is} symmetric, since $C_{ij}$ is actually
antisymmetric in $i$ and $j$). So one may safely integrate them out, after
performing a Wick rotation such as
\[
t\rightarrow \tau =it\quad ,\qquad \qquad C_{i10}\rightarrow \overline{C}
_{i10}=\pm iC_{i10}\quad .
\]
The  indeterminacy in the choice of the direction in which to perform the Wick rotation will turn out to be 
irrelevant after the integration of $C_{i 10}$ (indeed, this $\pm$ sign appears in each factor of $\phi$ and 
each factor of $G$, which always come in pairs).

We then get the Euclidean version of (\ref{actpert}):
\[
I_{\text{E}}\,=\,\gamma \int d\tau \,\left( \beta (\overline{\mathbf{C}}
_{i}^{a})^{\intercal }(\Id_{ij}^{ab}+\overline{\Delta }_{ij}^{ab})\overline{
\mathbf{C}}_{j}^{b}\,+\,(\overline{\mathbf{C}}_{i}^{a})^{\intercal }\overline{\mathbf{
F}}_{i}^{a}\,-\,\mathcal{L}_{C}\,-\,\mathcal{L}_{\phi }\,-\,\widehat{\mathcal{L}}\right) \quad ,
\]
where the new rotated fields assume the following form:
\begin{align*}
\qquad\qquad\overline{\mathbf{C}}_{i}^{a}\,& \triangleq &\,\left( 
\begin{array}{c}
C_{i0}^{a} \\ 
\overline{C}_{i10}^{a}
\end{array}
\right) 
\;,\qquad\qquad \qquad & \qquad
\overline{\mathbf{F}}_{i}^{a}\,\triangleq \,\left( 
\begin{array}{c}
-F_{i}^{a} \\ 
\pm iG_{i}^{a}
\end{array}
\right) \quad , 
\qquad\qquad\\
\qquad\qquad\Id_{ij}^{ab}\,&\triangleq &\,\left( 
\begin{array}{cc}
\delta ^{ab}\delta _{ij} & 0 \\ 
0 & \delta ^{ab}\delta _{ij}
\end{array}
\right) \;, \qquad& \qquad
\overline{\Delta }_{ij}^{ab}\, \triangleq \,\displaystyle{\frac{3f^{abc}}{\beta }}\left( 
\begin{array}{cc}
-C_{ij}^{c} & \pm i\phi ^{c}\delta _{ij} \\ 
\mp i\phi ^{c}\delta _{ij} & C_{ij}^{c}
\end{array}
\right) \quad .\qquad\qquad
\end{align*}
The gaussian integration is straightforward, and yields, after exponentiation of the
non trivial part of the determinant:
\begin{align*}
\int D\overline{C}_{i10}\,DC_{i0}\,\exp\Big\{-I_{\text{E}}\Big\}
\qquad\qquad\qquad\qquad\qquad\qquad\qquad\qquad\qquad\qquad\qquad
\qquad\qquad\qquad\qquad\qquad&\\
\qquad\propto\,
\exp\bigg\{-\frac{1}{2}Tr\left( \ln (\Id
_{ij}^{ab}+\overline{\Delta }_{ij}^{ab})\right) -\gamma \int d\tau \,\left( -
\frac{1}{4\beta }(\overline{\mathbf{F}}_{i}^{a})^{\intercal }(\Id_{ij}^{ab}+
\overline{\Delta }_{ij}^{ab})^{-1}\overline{\mathbf{F}}_{j}^{b}\,-\,\mathcal{
L}_{C}\,-\,\mathcal{L}_{\phi }\,-\,\widehat{\mathcal{L}}\right) \bigg\}\quad .
\end{align*}
The term quadratic in $\mathbf{F}$ is obviously tree-level, whereas the first one is a  1-loop correction
to the effective action. The 1-loop ''behaviour'' is encoded in the divergence associated with 
the trace of an operator, since
\begin{equation}\label{oper}
Tr\widehat{O}\,=\,\int d\tau\,O_{\phantom{i}i}^{i}(\tau)\langle \tau | \tau \rangle
\,=\,\Lambda \int d\tau \,O_{\phantom{i}i}^{i}(\tau)\quad ,
\end{equation}
where the integration in Fourier space is divergent, and has been replaced by the cutoff $\Lambda$.
Transforming back to real Minkowskian time $t$, we obtain the following effective 
action
\begin{equation}
I_{\text{eff}}\,=\,\gamma \int dt\,\left( \widehat{\mathcal{L}
}\,+\,\mathcal{L}_{C}\,+\,\mathcal{L}_{\phi }\,+\,\frac{1}{4\beta }(
\overline{\mathbf{F}}_{i}^{a})^{\intercal }(\Id_{ij}^{ab}+\overline{\Delta }
_{ij}^{ab})^{-1}\overline{\mathbf{F}}_{j}^{b} -\frac{\Lambda}{2\gamma} 
\big(\ln (\Id+\overline{\Delta}(t))\big)_{ii}^{aa}\right)\,.   \label{loop}
\end{equation}

\subsection{Analysis of the different contributions to the effective action}

The natural scale of (\ref{loop}) is $\beta $, which is proportional to the mass
parameter $\mu$. We therefore expand (\ref{loop}) in
powers of $1/\beta $, which amounts to expanding (\ref{loop})
in powers of $\overline{\Delta }$. Now, this procedure must be
regarded as a formal expansion, since we don't want to set $\beta$ to a particular value. 
However, this formal expansion in $1/\beta $ actually conceals a true expansion in $[X_{i},X_{j}]$, 
which should be small to minimize the potential energy, as will become clear later on.

First of all, let us consider the expansion of the tree-level term up to 
$\mathcal{O}(1/\beta ^{3})$. The first order term is given by: 
\begin{equation*}
\frac{1}{\beta }\int dt\,(\overline{\mathbf{F}}_{i}^{a})^{\intercal} \overline{\mathbf{F}}_{i}^{a}\,=
\,\frac{1}{\beta }\int
 dt\,Tr\Big((F_{i})^{2}-(G_{i})^{2}\Big) \,.
\end{equation*}
Since $F_{i}$ contains $[{\cal D}_t,X_i]$ and $\{\overline{\Psi},\Psi\}$, while $G_{i}$ 
contains only $\{\overline{\Psi},\Psi\}$ (ignoring $Z$-type contributions), this term will 
generate a kinetic term for the $X^i$'s as well as trilinear and quartic interactions.

The second-order term is:
\begin{equation*}
\frac{1}{\beta }\int dt\,(\overline{\mathbf{F}}_{i}^{a})^{\intercal }
\overline{\Delta }_{ij}^{ab}\overline{\mathbf{F}}_{j}^{b}\,=\,
\frac{3i}{\beta^{2}}\int dt\,Tr\bigg( C_{ij}\Big\{[F_{i},F_{j}]-[G_{i},G_{j}]\Big\}-2\,\phi\,[F_{i},G_{i}]\bigg) .
\end{equation*}
All vertices generated by this term contain either one $C$, with 2 to 4 $X$ or $\Psi$, or 
one $\phi$, with 3 or 4 $X$ or $\Psi$.

Finally, the third-order contibution is as follows:
\begin{align*}
\frac{1}{\beta }\int dt\,~(\overline{\mathbf{F}}_{i}^{a})^{\intercal }
(\overline{\Delta }^{2})_{ij}^{ab}\overline{\mathbf{F}}_{j}^{b}\,&=\,
-\frac{3^{2}}{\beta ^{3}}\int dt\,~Tr\bigg([F_{i},C_{ij}][C_{jk},F_{k}]\,-\,[G_{i},C_{ij}][C_{jk},G_{k}]\,+ \\
\,+[F_{i},\phi][\phi,F_{i}]\,&-\,[G_{i},\phi][\phi,G_{i}]\,+\,2\,[G_{i},C_{ij}][\phi,F_{j}]\,
-2\,[F_i,C_{ij}][\phi,G_{j}]\,\bigg)\, ,
\end{align*}
producing vertices with 2 $\phi$'s or 2 $C$'s, together with 2 to 4 $X$ or $\Psi$, as well as
vertices with 1 $\phi$ or 1 $C$, with 3 to 4 $X$ or $\Psi$.

Next we turn to the 1-loop term, where we expand the logarithm up to $\mathcal{O}(1/\beta ^{3})$.
Because of the total antisymmetry of both $f^{abc}$ and $C_{ij}$, one has $Tr\overline{\Delta }=0$, so that
the first term cancels. 
Now, keeping in mind that 
\[
f^{abc}f^{bad}\,=\,-C_{2}(\mathfrak{ad})\delta ^{cd}\text{ \qquad and \qquad }
f^{amn}f^{bno}f^{com}\,=\,\frac{1}{2}C_{2}(\mathfrak{ad})f^{abc}\quad ,
\]
$C_2 (\mathfrak{ad})$ referring to the quadratic Casimir operator in the adjoint representation of the Lie algebra,
one readily finds:
\renewcommand{\theenumi}{(\roman{enumi})}
\begin{enumerate}
\item  $Tr\overline{\Delta }^{2}\,=\,\left( \frac{3}{\beta }\right)
^{2}2iC_{2}(\mathfrak{ad})\Lambda \int dt \,Tr\Big( (C_{ij})^{2}-9(\phi )^{2}\Big) \,,$ 
\item  $Tr\overline{\Delta }^{3}\,=\,-\left( \frac{3}{\beta }\right)
^{3}C_{2}(\mathfrak{ad})\Lambda \int dt \,Tr\Big( C_{ij}[C_{jk},C_{ki}]\Big) \,. $
\end{enumerate}
In other words, the 1-loop correction (i) renormalizes the mass terms
for $C_{ij}$ and $\phi $ in $\widetilde{I}_{c}$ as follows:
\begin{itemize}
\item  Mass renormalization for $C_{ij}$: $\frac{1}{2}\gamma \beta
\longrightarrow \frac{1}{2}\gamma \beta \left( 1+\frac{3^{2}}{\gamma\beta ^{3}}
C_{2}(\mathfrak{ad})\Lambda \right) $
\item  Mass renormalization for $\phi $: $\gamma\beta \longrightarrow \gamma\beta \left(
1+\frac{3^{4}}{2\gamma\beta^{3}}C_{2}(\mathfrak{ad})\Lambda \right) $
\end{itemize}
Whereas the 1-loop correction (ii) renormalizes the trilinear coupling
between the $C_{ij}$ in $I_{c}$:
\begin{itemize}
\item  Renormalization of the $C_{ij}[C_{jk},C_{ki}]$ coupling: $
\gamma\longrightarrow \gamma\left( 1-\frac{3^2}{2\gamma\beta^3}C_{2}(\mathfrak{ad})\Lambda \right)$
\end{itemize}

Up to $Tr\overline{\Delta }^{3}$, the 1-loop corrections actually only renormalize terms already present in 
$I_{c}$ from the start. This is not the case for the higher order subsequent 1-loop corrections: 
there is an infinite number of such corrections, each one diverging like $\Lambda$. 
A full quantization of (\ref{loop}) is obviously a formidable task, which we will not attempt
in the present paper. A sensible regularization of the divergent contributions should take into account
the symmetries of the classical action, which are not explicit anymore after performing T-dualities and
the IMF limit. However, since our model is quantum-mechanical, we believe it to be finite even if we haven't
come up with a fully quantized formulation. 

Summing up the different contributions computed in this section, one gets the following 1-loop effective action 
up to $\mathcal{O}(1/\beta ^{3})$:
\begin{eqnarray}
\frac{1}{\gamma} &I_{\text{eff}}\,&=\,\int dt\,\left( \mathcal{L}_{C}\,+\,\mathcal{L}_{\phi }\,+
\,\widehat{\mathcal{L}}\right) \,+\,\frac{\gamma}{4\beta }\int dt\,
\,Tr\left(F_{i}^{2}-G_{i}^{2}\right)\,-  \nonumber \\
&&-\,\frac{3i\gamma}{4\beta ^{2}}\int dt\,Tr\bigg(C_{ij}\Big([F_{i},F_{j}]-[G_{i},G_{j}]\Big)-2\,\phi\,[F_{i},G_{i}]\bigg)
+\,\frac{\gamma\lambda}{2\beta^2}\int dt\,Tr\bigg(C_{ij}^{2}-9\phi ^{2}\bigg)\,-\nonumber\\
&&-\,\frac{9\gamma}{4\beta ^{3}}\int dt\,Tr\bigg([F_{i},C_{ij}][C_{jk},F_{k}]\,-\,[G_{i},C_{ij}][C_{jk},G_{k}]\,+\, 
[F_{i},\phi][\phi,F_{i}]\,-\,[G_{i},\phi][\phi,G_{i}]\,+\nonumber\\
&&+\,2\,[G_{i},C_{ij}][\phi,F_{j}]\,-2\,[F_i,C_{ij}][\phi,G_{j}]\bigg)-\,
\frac{i\lambda\gamma}{2\beta^3}\int dt\,Tr\bigg( \,C_{ij}\,[C_{jk},C_{ki}]\bigg) \,+\,\mathcal{O}(1/\beta ^{4})\quad .
\label{loop2}
\end{eqnarray}
where $\lambda$ is proportional to the cutoff $\Lambda$: 
\[
\lambda\,\triangleq \,\frac{9\,C_{2}(\mathfrak{ad})\Lambda }{\gamma }\quad .
\]
Note that the $\mathcal{O}(1/\beta ^{4})$ terms that we haven't written contain at least three powers of $C_{ij}$ or 
$\phi$. 

\subsection{Iterative solution of the constraint equations}

The 1-loop corrected action (\ref{loop2}) still contains the constraint fields 
$C_{ij}$ and $\phi$, which should in principle be integrated out in order to get
the final form of the effective action. Since $I_{\text{eff}}$ contains arbitrarily 
high powers of $C_{ij}$ and $\phi$, we cannot perform a full path integration. 
We can however solve the equations for $C_{ij}$ and $\phi$ perturbatively in $1/\beta $. 
This allows to replace these fields in (\ref{loop2}) with the solution to their equations of motion. 
Thus, in contrast with the preceeding subsection, here we remain at tree-level.

The equation of motion for $C_{ij}$ may be
computed from (\ref{loop2}), and reads:
\begin{eqnarray}
&&C_{ij}\,+\,\frac{1}{\beta }\Big( E_{ij}+3i[C_{jk},C_{ki}]\Big) \,+\,
\frac{1}{\beta ^{3}}\left( \frac{3}{4}i\Big\{
[G_{i},G_{j}]-[F_{i},F_{j}]\Big\}+\lambda C_{ij}\right) \,+  \nonumber \\
&&\quad +\,\frac{1}{\beta ^{4}}\, \frac{9}{2}\left(\Big\{[F_{[i},[C_{j]k},F_{k}]]-[G_{[i},[C_{j]k},G_{k}]]+
[G_{[i},[\phi,F_{j]}]]-[F_{[i},[\phi ,G_{j]}]]\Big\} \right.\,+\nonumber\\
&&\quad-\,\left.\frac{i\lambda}{3}[C_{jk},C_{ki}]\right)  
\,+\,\mathcal{O}(1/\beta ^{5})\,=\,0\;,  \label{eom1}
\end{eqnarray}
while the equation of motion for $\phi$ is:
\begin{eqnarray}
&&\phi \,-\,\,\frac{1}{2\beta }J\,-\,\frac{3}{\beta ^{3}}\left( \frac{i}{4}\,[F_{i},G_{i}]-
3\lambda \phi \right) \,+\,
\frac{3^2}{4\beta^{4}}\bigg([F_{i},[F_{i},\phi]]\,- \nonumber \\
&&\qquad \qquad -\,[G_{i},[G_{i},\phi]]\,+\,[F_{i},[C_{ij},G_{j}]]\,-\,[G_{i},[C_{ij},F_{j}]]\bigg) \,+\,
\mathcal{O}(1/\beta ^{5})\,=\,0\;.  \label{eom2}
\end{eqnarray}
By solving the coupled equations of motion (\ref{eom1}) and (\ref{eom2}) recursively, 
one gets $C_{ij}$ and $\phi $ up to $\mathcal{O}(1/\beta ^{5})$. 
We can safely stop at $\mathcal{O}(1/\beta ^{5})$, because the terms contributing to that order
in~(\ref{eom1}) and~(\ref{eom2}) are, on the one hand, 
$\beta^{-1}\Lambda (\delta /\delta C_{ij})Tr\overline{\Delta }^4$ and 
$\beta^{-1}\Lambda (\delta /\delta \phi )Tr\overline{\Delta }^4$, whose lowest order
is $\mathcal{O}(1/\beta ^{8})$, and on the other hand $\beta ^{-2}(\delta
/\delta C_{ij})\mathbf{F}^{\intercal }\overline{\Delta }^3\mathbf{F}$ and $
\beta ^{-2}(\delta /\delta \phi )\mathbf{F}^{\intercal }\overline{\Delta }^3
\mathbf{F}$, whose lowest order is $\mathcal{O}(1/\beta ^{7})$, so that the eom don't get 
any corrections from contributions of $\mathcal{O}(1/\beta ^{4})$ coming from $I_{\text{eff}}$.

Subsequently, the $1/\beta $ expansion for $C_{ij}$ reads
\begin{align}\label{eo1}
C_{ij}\, =&\,-\frac{1}{\beta }E_{ij}\,+\,\frac{3i}{\beta ^{3}}\left(
[E_{ik},E_{kj}]+\frac{1}{4}[F_{i},F_{j}]-\frac{1}{4}[G_{i},G_{j}]\right)
\,+\,\frac{\lambda}{\beta ^{4}}E_{ij}\,+\nonumber\\
&\,+\,\frac{9}{\beta ^{5}}\bigg(-2[E_{[ik},[E_{kl},E_{lj]}]] \,+ \,
\frac{1}{2}[E_{[ik},[F_{k},F_{j]}]]\,-\,\frac{1}{2}[E_{[ik},[G_{k},G_{j]}]]\,+\,
\frac{1}{2}[[E_{[ik},F_{k}],F_{j]}]- \nonumber\\
&\,-\,\frac{1}{2}[[E_{[ik},G_{k}],G_{j]}]\,+\,\frac{1}{4}[G_{[i},[F_{j]},J]]-
\frac{1}{4}[F_{[i},[G_{j]},J]]\bigg) \,+\,\mathcal{O}(1/\beta ^{6})\quad ,
\end{align}
and the expansion for $\phi $: 
\begin{eqnarray}\label{eo2}
\phi \, &=&\,\frac{1}{2\beta }J\,+\,\frac{3i}{4\beta ^{3}}[F_{i},G_{i}]\,-
\,\left( \frac{3}{2}\right) ^{2}\frac{\lambda}{\beta ^{4}}J\,-\,\\
&&-\,\frac{9}{8\beta ^{5}}\bigg( [F_{i},[F_{i},J]]-[G_{i},[G_{i},J]]\,- 
\,2[[F_{i},E_{ij}],G_{j}]+2[[G_{i},E_{ij}],F_{j}]\bigg) +\mathcal{
O}(1/\beta ^{6})\;.\nonumber
\end{eqnarray}

Now, plugging the result for $C_{ij}$ and $\phi $ into $I_{\text{eff}}$, one
arrives at the ''perturbative'' effective action, which we have written
up to and including $\mathcal{O}(1/\beta ^{5})$, since the highest order ($\mathcal{O}(1/\beta ^{3})$)
we calculated in $I_{\text{eff}}$ is quadratic in $C$ and $\phi$\footnote{note that their expansion starts at 
$\mathcal{O}(1/\beta)$}, and since the $\mathcal{O}(1/\beta ^{4})$-terms in~(\ref{loop2}) only generate 
$\mathcal{O}(1/\beta ^{7})$~-~terms. This effective action takes the following form:
\begin{eqnarray*}
\frac{1}{\gamma}\,I_{\text{eff}} &=&\int dt\,\Bigg( \widehat{\mathcal{L}}+
\frac{1}{4\beta}\,Tr\left(F_{i}^{2}-G_{i}^{2}+J^{2}-2\,(E_{ij})^2\right) + \\
&&+\frac{i}{\beta ^{3}}\,Tr\bigg(-E_{ij}[E_{jk},E_{ki}]\,+\,\frac{3}{4}E_{ij}\Big\{[F_{i},F_{j}]-[G_{i},G_{j}]\Big\}+
\frac{3}{4}J[F_{i},G_{i}]\bigg) \Bigg)
 +\\
&&+\frac{\lambda}{2\beta ^{4}}\,Tr\bigg((E_{ij})^2-\frac{9}{4}J^2\bigg)+
\frac{9}{2\beta ^{5}}\,Tr\bigg(([E_{ik},E_{kj}])^2 + \frac{1}{16}([F_i,F_j]-[G_i,G_j])^2-\\
&&+\frac{1}{2}[E_{ik},E_{kj}]\big([F_i,F_j]-[G_i,G_j]\big)-
\frac{1}{8}([F_i,G_i])^2-\frac{1}{2}\Big\{([F_i,E_{ij}])^2-([G_i,E_{ij}])^2\Big\}+\\
&&+\frac{1}{4}\Big\{([F_i,J])^2-([G_i,J])^2\Big\}-\frac{1}{2}[G_i,E_{ij}][J,F_j]+\frac{1}{2}[F_i,E_{ij}][J,G_j]\bigg)\Bigg)
+{\cal O}(1/\beta^6) \,.
\end{eqnarray*}

At that point, we can replace the aliases $E$, $F$, $G$ and $J$ by their expression in terms of
the fundamental fields $X$, $\Psi$, $Z$, $\Pi$, $B$ and $H$. The result of this lengthy computation
(already to order $1/\beta$) is presented in the Appendix.
Here, we will only display the somewhat simpler result obtained by ignoring all 5-form induced fields.
Furthermore, we will remove the parameter $\beta$ from the action, since it was only 
useful as a reminder of the order of calculation in the perturbative approach. To do so, 
we absorb a factor of $1/\beta$ in every field, as well as in ${\mathcal D}_t$ (so that 
the measure of integration scales with $\beta$).
Thus, $\beta$ only appears in the prefactor in front of the action, at the 4$^{th}$ power.
This is similar to the case of Yang-Mills theory, where one can choose either to have
a factor of the coupling constant in the covariant derivatives or have it as a prefactor
in front of the action. To be more precise, we set:
\begin{equation*}
\Theta \,= \,\frac{1}{4\sqrt{6}\beta} \,\Psi\,,\quad \widetilde{X}_i\,=\,\frac{1}{\beta}\,X_i\,,
\quad \widetilde{A}_0\,=\,\frac{1}{\beta}\,A_0\,,\quad G\,=\,9 \beta^4\, \gamma\,,\quad \tilde{t}\,=\,
\beta t\;,
\end{equation*}
and similarily for the $Z$ sector: $(Z,\Pi,H,B)\rightarrow (Z/\beta ,\Pi/\beta ,H/\beta ,B/\beta )$.

With this redefinition, it becomes clear that our developpment is really an expansion in 
higher commutators and not in $\beta$. It makes thus sense to limit it to the lowest orders 
since the commutators should remain small to minimize the potential energy.
To get a clearer picture of the final result, we will put all the 5-form-induced fields $(Z,\Pi,H,B)$ to zero.
For convenience we will still write $\widetilde{X}$ as $X$ and $\tilde{t}$ as $t$ in 
the final result, which reads:
\begin{align*}
I(X,\Theta)=& \frac{1}{G}\int dt\, Tr_{\mathfrak{u}(N)} \Bigg(([{\cal D}_t,X_i])^2 + \frac{1}{2}([X_i,X_j])^2 +
i\overline{\Theta}\tilde{\G}_0 [{\cal D}_t,\Theta]-
\overline{\Theta}\tilde{\G}_i [X_i,\Theta] - \\
-&\frac{1}{9} (X_i)^2 - \frac{2i}{3}\overline{\Theta}\Theta - 
3 [{\cal D}_t,X_i] \{\overline{\Theta},\tilde{\G}_i \tilde{\G}_0 \Theta\} - 
\frac{3i}{2}[X_i,X_j]\{\overline{\Theta},\tilde{\G}_{ij} \Theta\} + \\
+&\frac{9}{4} (\{\overline{\Theta},\tilde{\G}_i \tilde{\G}_0 \Theta\})^2 - 
\frac{9}{4} (\{\overline{\Theta},\tilde{\G}_i \tilde{\G}_* \Theta\})^2 + 
\frac{9}{4} (\{\overline{\Theta},\tilde{\G}_0 \tilde{\G}_* \Theta\})^2 - 
\frac{9}{8} (\{\overline{\Theta},\tilde{\G}_{ij} \Theta\})^2 + \\
+&3 [X_i,X_j][[X_j,X_k],[X_k,X_i]] - 9 [X_i,X_j][[{\cal D}_t,X_i],[{\cal D}_t,X_j]]- \\
-& \frac{3^3i}{2} \{\overline{\Theta},\tilde{\G}_{ij} \Theta\} [[X_j,X_k],[X_k,X_i]] + 
\frac{3^4}{2^2} [X_i,X_j]
[\{\overline{\Theta},\tilde{\G}_{jk} \Theta\},\{\overline{\Theta},\tilde{\G}_{ki} \Theta\}] - \\
-& \frac{3^4i}{2^3}\{\overline{\Theta},\tilde{\G}_{ij} \Theta\}
[\{\overline{\Theta},\tilde{\G}_{jk} \Theta\},\{\overline{\Theta},\tilde{\G}_{ki} \Theta\}]+ 
\frac{3^3i}{2}\{\overline{\Theta},\tilde{\G}_{ij} \Theta\}[[{\cal D}_t,X_i],[{\cal D}_t,X_j]] + \\
+& 3^3 [X_i,X_j][[{\cal D}_t,X_i],\{\overline{\Theta},\tilde{\G}_j \tilde{\G}_0 \Theta\}] -
\frac{3^4i}{2^2}\{\overline{\Theta},\tilde{\G}_{ij} \Theta\}
[[{\cal D}_t,X_i],\{\overline{\Theta},\tilde{\G}_j \tilde{\G}_0 \Theta\}] - \\
-& \frac{3^4}{2^2}[X_i,X_j]
[\{\overline{\Theta},\tilde{\G}_i \tilde{\G}_0 \Theta\},\{\overline{\Theta},\tilde{\G}_j \tilde{\G}_0 \Theta\}] +
\frac{3^5i}{2^3}\{\overline{\Theta},\tilde{\G}_{ij} \Theta\}
[\{\overline{\Theta},\tilde{\G}_i \tilde{\G}_0 \Theta\},\{\overline{\Theta},\tilde{\G}_j \tilde{\G}_0 \Theta\}]+\\
+& \frac{3^4}{2^2}[X_i,X_j]
[\{\overline{\Theta},\tilde{\G}_i \tilde{\G}_* \Theta\},\{\overline{\Theta},\tilde{\G}_j \tilde{\G}_* \Theta\}] - 
\frac{3^5i}{2^3}\{\overline{\Theta},\tilde{\G}_{ij} \Theta\}
[\{\overline{\Theta},\tilde{\G}_i \tilde{\G}_* \Theta\},\{\overline{\Theta},\tilde{\G}_j \tilde{\G}_* \Theta\}]-\\
-& \frac{3^4i}{2} \{\overline{\Theta},\tilde{\G}_0 \tilde{\G}_* \Theta\}
[[{\cal D}_t,X_i],\{\overline{\Theta},\tilde{\G}_i \tilde{\G}_* \Theta\}] +,
\frac{3^5i}{2}\{\overline{\Theta},\tilde{\G}_0 \tilde{\G}_* \Theta\}[\{\overline{\Theta},\tilde{\G}_i \tilde{\G}_0 \Theta\},\{\overline{\Theta},\tilde{\G}_i \tilde{\G}_* \Theta\}] \Bigg) + \\
+&\,\text{eighth-order interactions.}
\end{align*}
We see that the first four terms in this action correspond to the BFSS matrix model, but with a doubled
number of fermions. So, in order to maintain half of the original supersymmetries (i.e. ${\cal N}=1$ in $10D$), one could
project out half of the original fermions with ${\cal P}_{-}\xrightarrow{\text{IMF}}(1 + \tilde{\G}_{*})/2$. 
Finally, in addition to the BFSS-like terms, we have mass terms and an infinite tower of interactions possibly containing
information about the behaviour of brane dynamics in the non-perturbative sector.

\section{Discussion}

After a general description of $\mathfrak{osp}(1|32)$ and its adjoint representation, we have studied its expression as
a symmetry algebra in 12$D$. We have described the resulting transformations of matrix fields and their commutation 
relations. Finally, we have proposed a matrix theory action possessing this symmetry in 12$D$. We have then repeated
this analysis in the 11-dimensional case, where $\mathfrak{osp}(1|32)$ is a sort of super-$AdS$ algebra.
Compactification and T-dualization of two coordinates produced a one-parameter family of singular limiting procedures 
that shrink the world-sheet along a world-line. We have then identified one of them as the usual IMF limit, which 
gave rise to a non-compact dynamical evolution parameter that has allowed us to distinguish dynamical from auxiliary
fields. Integrating out the latter and solving some constraints recursively, we have obtained a matrix
model with a highly non-trivial dynamics, which is similar to the BFSS matrix model when both $X^{2}$ and multiple
commutators are small. The restriction to a low-energy sector where both $X^{2}$ and $[X,X]$ are small seems 
to correspond to a space-time with weakly interacting (small $[X,X]$) D-particles that are nevertheless not far apart
(small $X^2$). The stable classical solutions correspond to vanishing matrices, i.e. to D-particles stacked at 
the origin, which diplays some common features with matrix models in pp-wave backgrounds (see for 
instance~\cite{BMN,DSJvR,Bon}).

Since the promotion of the membrane charges in the $11D$ super-Poincar\'e algebra to symmetry generators
implied the non-commutativity of the $P$'s, and thus the $AdS_{11}$ symmetry, the membranes are responsible for
some background curvature of the space-time. Indeed, since the $C_{MN}$ don't appear as dynamical degrees of freedom, 
their r\^ole is to produce the precise tower of higher-order interactions necessary to enforce such a global symmetry
on the space-time dynamically generated by the $X_i$'s.
The presence of mass terms is thus no surprise since they were also conjectured to appear in matrix models aimed 
at describing gravity in deSitter spaces, albeit with a tachyonic sign reflecting the unusual causal structure of 
deSitter space~(\cite{Gao,Cham}).
One might also wonder whether the higher interaction terms we get are somehow related to the high energy corrections 
to BFSS one would obtain from the non-abelian Dirac-Born-Infeld action. Another question one could address is 
what kind of corrections a term of the form $STr_{\mathfrak{osp}(1|32)\otimes \mathfrak{u}(n)} ([M,M][M,M])$ would 
induce.

It would also be interesting to investigate the dynamics of the 5-branes degrees of freedom more thoroughly by computing
the effective action for $Z$ (from $I_{\text{eff}}$ of the Appendix) and give a definite proposal for the physics of 
5-branes in M-theory. Note that there is some controversy about the ability of the BFFS model to describe transverse 
5-branes (see e.g. \cite{GRWati,BSS} and references therein for details). Our model would provide an interesting 
extension of the BFSS theory by introducing in a very natural way transverse 5-branes (through the fields dual to 
$Z_{ijkl}$) in addition to the D0-branes bound states describing longitudinal 5-branes, which are already present 
in BFSS theory. 

\section{Acknowledgements}
The authors want to thank J.-P. Derendinger, C.-S. Chu, J. Walcher, V. Braun, C. R\"omelsberger, M. Cederwall, 
L. Smolin, D. Buchholz and F. Ferrari for useful discussions during the preparation of this work, as well as
W. Taylor IV, R. Helling, J.Plefka and I. Ya. Aref'eva for comments preceding the revised version. M. B. warmly 
thanks H. Kawai, T. Yokono, I. Ojima and everyone else at Kyoto University for the opportunity of presenting 
this work there and their hospitality during his stay in Kyoto. The authors acknowledge financial support from the Swiss Office 
for Education and Science, the Swiss National Science Foundation and the European Community's Human Potential
Programme.

\section{Appendix}

We give here the complete effective action at order $1/\beta$.
\begin{eqnarray*}\label{BFSS}
I_{\text{eff}} &=&\frac{1}{G}\,\int dt\,Tr_{\frak{u}(N)}\,
\Bigg(-\beta\bigg\{(X_{i})^2+\frac{i}{16}\,\overline{\Psi }\Psi+
\frac{1}{4!}\left(\left(Z_{i_{1}\cdots i_{4}}\right)^{2}+
\left(\Pi _{i_{1}\cdots i_{4}}\right)^{2}-
\left(H_{i_{1}\cdots i_{4}}\right)^{2}-
4\left( B_{i_{1}i_{2}i_{3}}\right)^{2}\right)\bigg\}+\nonumber\\
&+&\bigg\{\frac{1}{4}\Pi_{i_{1}\cdots i_{4}}[\mathcal{D}_{t},Z_{i_{1}\cdots i_{4}}]+
\frac{3i}{32}\overline{\Psi }\widetilde{\Gamma }_{0}[\mathcal{D}_{t},\Psi]+
i\Pi_{i_{1}i_{2}i_{3}\,j}[X_{j},B_{i_{1}i_{2}i_{3}}]-
\frac{i}{4\cdot4!}\varepsilon_{i_{1}\cdots i_{8} j}Z_{i_{1}\cdots i_{4}}[X_{j},H_{i_{5}\cdots i_{8}}]+\nonumber\\
&+&\frac{i}{3!\cdot 4!}\varepsilon_{i_{1}\cdots i_{9}}\,\bigg(
B_{i_{1}i_{2}\,j}\,\Big(2\,[\Pi_{j\,i_{3}i_{4}i_{5}},\Pi_{i_{6}\cdots i_{9}}]\,
+[Z_{j\,i_{3}i_{4}i_{5}},Z_{i_{6}\cdots i_{9}}]\,
-[H_{j\,i_{3}i_{4}i_{5}},H_{i_{6}\cdots i_{9}}]\Big)+\nonumber\\
&+&\frac{2}{3}\,B_{i_{1}i_{2}i_{3}}\,\Big(\,[B_{i_{4}i_{5}i_{6}},B_{i_{7}i_{8}i_{9}}]\,
+\,[Z_{i_{4}i_{5}i_{6}\,j},Z_{j\,i_{7}i_{8}i_{9}}]\,
-\,[H_{i_{4}i_{5}i_{6}\,j},H_{j\,i_{7}i_{8}i_{9}}]\,\Big)\bigg)\,+\nonumber\\
&+&\frac{i}{4}\,\Pi_{i_{1}i_{2}j_{1}j_{2}}\,[Z_{j_{1}j_{2}k_{1}k_{2}},H_{k_{1}k_{2}i_{1}i_{2}}]\,-\,
\frac{3}{32}\,\bigg( \overline{\Psi }\,\widetilde{\Gamma }_{i}\,[X_{i},\Psi ]\,+\,
\frac{1}{4!}\,\overline{\Psi }\,\left(\widetilde{\Gamma}_{i_1\cdots i_4}\widetilde{\Gamma}_{*}
[Z_{i_1\cdots i_4},\Psi ]\,+\right.\nonumber\\
&+&\left.\,\widetilde{\Gamma}_{i_1\cdots i_4}\widetilde{\Gamma}_{0}\widetilde{\Gamma}_{*}[\Pi_{i_1\cdots i_4},\Psi ]
\,-\,
\widetilde{\Gamma}_{i_1\cdots i_4}\widetilde{\Gamma}_{0}[H_{i_1\cdots i_4},\Psi ]\,-\,
4\,\widetilde{\Gamma}_{i_1 i_2 i_3}\widetilde{\Gamma}_{0}\widetilde{\Gamma}_{*}[B_{i_1 i_2 i_3},\Psi]\right)\bigg)
\bigg\}\,+ 
\nonumber \\
+&\displaystyle{\frac{1}{4 \beta}}&\,\bigg\{36\,([{\cal D}_{t},X_{i}])^{2}\,-\,
\frac{i}{8}\varepsilon_{ij_1 \cdots j_8}[{\cal D}_{t},X_{i}][H_{j_1 \cdots j_4},\Pi_{j_5 \cdots j_8}]\,-\,
12i\,[{\cal D}_{t},X_{i}][Z_{i j_1 \cdots j_3},B_{j_1 \cdots j_3}]\,-\nonumber\\
&-&\frac{9}{8}[{\cal D}_{t},X_{i}]\{\overline{\Psi },\widetilde{\Gamma }_{i}\widetilde{\Gamma }_{0}\Psi\}\,-\,
\frac{1}{16}[H_{i_1 \cdots i_4},\Pi_{j_1 \cdots j_4}]\Big([H_{i_1 \cdots i_4},\Pi_{j_1 \cdots j_4}]\,-\,
16[H_{i_1 i_2 i_3 j_4},\Pi_{j_1 j_2 j_3 i_4}]\,+\nonumber \\
&+&\,36[H_{i_1 i_2 j_3 j_4},\Pi_{j_1 j_2 i_3 i_4}]\,-\,
16[H_{i_1 j_2 j_3 j_4},\Pi_{j_1 i_2 i_3 i_4}]\,+\,[H_{j_1 j_2 j_3 j_4},\Pi_{i_1 i_2 i_3 i_4}]\Big)\,-\nonumber\\
&-&\frac{1}{2\cdot 4!}\varepsilon_{ij_1 \cdots j_8}[H_{j_1 \cdots j_4},\Pi_{j_5 \cdots j_8}][Z_{i k_1 \cdots k_3},B_{k_1 \cdots k_3}]\,+\,
\frac{i}{2^9}\varepsilon_{ij_1 \cdots j_8}[H_{j_1 \cdots j_4},\Pi_{j_5 \cdots j_8}]\{\overline{\Psi },\widetilde{\Gamma }_{i}\widetilde{\Gamma }_{0}\Psi\}\,-\nonumber\\
&-&([Z_{i j_1 \cdots j_3},B_{j_1 \cdots j_3}])^2\,+\,
\frac{3i}{16}[Z_{i j_1 \cdots j_3},B_{j_1 \cdots j_3}]\{\overline{\Psi },\widetilde{\Gamma }_{i}\widetilde{\Gamma }_{0}\Psi\}\,+\,\frac{9}{2^{10}}(\{\overline{\Psi },\widetilde{\Gamma }_{i}\widetilde{\Gamma }_{0}\Psi\})^2\,+\nonumber\\
&+&\frac{1}{16}\,[Z_{i_1 \cdots i_4},\Pi_{j_1 \cdots j_4}]\Big([Z_{i_1 \cdots i_4},\Pi_{j_1 \cdots j_4}]\,-\,
16[Z_{i_1 i_2 i_3 j_4},\Pi_{j_1 j_2 j_3 i_4}]\,+\,36[Z_{i_1 i_2 j_3 j_4},\Pi_{j_1 j_2 i_3 i_4}]\,-\nonumber\\
&-&\,16[Z_{i_1 j_2 j_3 j_4},\Pi_{j_1 i_2 i_3 i_4}]\,+\,[Z_{j_1 j_2 j_3 j_4},\Pi_{i_1 i_2 i_3 i_4}]\Big)\,-\,
\frac{1}{2\cdot 4!}\varepsilon_{ij_1 \cdots j_8}[Z_{j_1 \cdots j_4},\Pi_{j_5 \cdots j_8}][H_{i k_1 \cdots k_3},B_{k_1 \cdots k_3}]\,-\nonumber\\
&-&\frac{i}{2^9}\varepsilon_{ij_1 \cdots j_8}[Z_{j_1 \cdots j_4},\Pi_{j_5 \cdots j_8}]\{\overline{\Psi },\widetilde{\Gamma }_{i}\widetilde{\Gamma }_{*}\Psi\}\,+\,([H_{i j_1 \cdots j_3},B_{j_1 \cdots j_3}])^2\,+\,
\frac{3i}{16}[H_{i j_1 \cdots j_3},B_{j_1 \cdots j_3}]\{\overline{\Psi },\widetilde{\Gamma }_{i}\widetilde{\Gamma }_{*}\Psi\}\,-\nonumber\\
&-&\frac{9}{2^{10}}(\{\overline{\Psi },\widetilde{\Gamma }_{i}\widetilde{\Gamma }_{*}\Psi\})^2-
\frac{1}{16}([Z_{i_1 \cdots i_4},H_{i_4 \cdots i_4}])^2+
\frac{3i}{2^6}[Z_{i_1 \cdots i_4},H_{i_4 \cdots i_4}]\{\overline{\Psi },\widetilde{\Gamma }_{0}\widetilde{\Gamma }_{*}\Psi\}+\frac{9}{2}([B_{i k_1 k_2},B_{j k_1 k_2}])^2+\nonumber\\
&+&\frac{1}{2}([Z_{i k_1 k_2 k_3},Z_{j k_1 k_2 k_3}])^2\,+\,                            
[Z_{i k_1 k_2 k_3},Z_{j k_1 k_2 k_3}][\Pi_{i l_1 l_2 l_3},\Pi_{j l_1 l_2 l_3}]\,+\,            
\frac{1}{2}([\Pi_{i k_1 k_2 k_3},\Pi_{j k_1 k_2 k_3}])^2\,-\nonumber\\
&-&3[Z_{i k_1 k_2 k_3},Z_{j k_1 k_2 k_3}][B_{i l_1 l_2},B_{j l_1 l_2}]\,-\,
[Z_{i k_1 k_2 k_3},Z_{j k_1 k_2 k_3}][H_{i l_1 l_2 l_3},H_{j l_1 l_2 l_3}]\,+\,
\frac{9}{2^{10}}(\{\overline{\Psi },\widetilde{\Gamma }_{0}\widetilde{\Gamma }_{*}\Psi\})^2\,-\nonumber\\
&-&3\,[\Pi_{i k_1 k_2 k_3},\Pi_{j k_1 k_2 k_3}][B_{i l_1 l_2},B_{j l_1 l_2}]\,-\,
[\Pi_{i k_1 k_2 k_3},\Pi_{j k_1 k_2 k_3}][H_{i l_1 l_2 l_3},H_{j l_1 l_2 l_3}]\,+\nonumber\\
&+&3\,[B_{i k_1 k_2},B_{j k_1 k_2}][H_{i l_1 l_2 l_3},H_{j l_1 l_2 l_3}]\,+\,
\frac{1}{2}([H_{i k_1 k_2 k_3},H_{j k_1 k_2 k_3}])^2\,-\,
6\,[Z_{i k_1 k_2 k_3},Z_{j k_1 k_2 k_3}][X_{i},X_{j}]\,+\nonumber\\
\end{eqnarray*}

\begin{eqnarray}
&+&\frac{3i}{32}[Z_{i k_1 k_2 k_3},Z_{j k_1 k_2 k_3}]\{\overline{\Psi },\widetilde{\Gamma }_{ij}\Psi\}-
6[\Pi_{i k_1 k_2 k_3},\Pi_{j k_1 k_2 k_3}][X_{i},X_{j}]+
\frac{3i}{32}[\Pi_{i k_1 k_2 k_3},\Pi_{j k_1 k_2 k_3}]\{\overline{\Psi },\widetilde{\Gamma }_{ij}\Psi\}+\nonumber\\
&+&18\,[B_{i k_1 k_2},B_{j k_1 k_2}][X_{i},X_{j}]\,-\,
\frac{9i}{32}[B_{i k_1 k_2},B_{j k_1 k_2}]\{\overline{\Psi },\widetilde{\Gamma }_{ij}\Psi\}\,+\,
6\,[H_{i k_1 k_2 k_3},H_{j k_1 k_2 k_3}][X_{i},X_{j}]\,-\nonumber\\
&-&\frac{3i}{32}[H_{i k_1 k_2 k_3},H_{j k_1 k_2 k_3}]\{\overline{\Psi },\widetilde{\Gamma }_{ij}\Psi\}+
18\,([X_{i},X_{j}])^{2}\,-\,\frac{9i}{16}[X_{i},X_{j}]\{\overline{\Psi },\widetilde{\Gamma }_{ij}\Psi\}-
\frac{9}{2^{11}}(\{\overline{\Psi},\widetilde{\Gamma }_{ij}\Psi\})^2\,\bigg\}+\nonumber\\
&+& {\cal O}(1/\beta^3) \Bigg)\,. \nonumber
\end{eqnarray}

\end{document}